\documentclass[11pt]{article}
\usepackage{mathrsfs}
\usepackage{latexsym}
\usepackage{amsmath,amssymb,amsfonts, amsthm}
\usepackage[pdftex]{hyperref}
\usepackage{amsthm}
\usepackage{amsfonts,cite}
\usepackage[usenames]{color}
\usepackage{amssymb}
\usepackage{graphicx}
\usepackage{amsmath}
\usepackage{amsfonts}
\usepackage{amsthm}
\usepackage{mathrsfs}
\usepackage{dsfont}
\newcommand\be{\begin{equation}}
\newcommand\ee{\end{equation}}
\newcommand\ber{\begin{eqnarray}}
\newcommand\eer{\end{eqnarray}}
\newcommand\berr{\begin{eqnarray*}}
\newcommand\eerr{\end{eqnarray*}}
\newcommand\bea{\begin{eqnarray}}
\newcommand\eea{\end{eqnarray}}
\newcommand\ba{\begin{array}}
\newcommand\ea{\end{array}}
\newcommand\bfR{\mathbb{R}}
\newcommand\dd{\mathrm{d}}
\newcommand\lm{\lambda}
\newcommand\ii{\mbox{i}}
\newcommand\e{\mathrm{e}}
\newcommand\eq{\eqref}\newcommand\lb{\label}

\newcommand\ri{\mathrm{i}}
\newcommand\ka{\kappa}
\newcommand\pa{\partial}

\newcommand{\nn}{\nonumber}

\newcommand\Om{\Omega}

\newcommand{\vep}{\varepsilon}
\newcommand{\bi}{\begin{itemize} }
  \newcommand{\ei}{\end{itemize} }

\newtheoremstyle{mythm}{1.5ex plus 1ex minus .2ex}{1.5ex plus 1ex
minus .2ex}{\kai}{\parindent}{\song\bfseries}{}{1em}{}
\numberwithin{equation}{section}\numberwithin{figure}{section}

\allowdisplaybreaks[4]
\begin{document}
\title{Bogomol'nyi Equations in  Mixed Product Chern--Simons \\ Theories Governing Charged Vortices and  Antivortices}
\author{Aonan Xu \\ School of Mathematics and Statistics\\Henan University\\
Kaifeng, Henan 475004, P. R.  China}
\date{}
\maketitle

\begin{abstract}
We extend product Chern-Simons theory to develop several mixed $U(1)\times U(1)$ models where one gauge field is governed by a Chern-Simons term and the other by a Maxwell or Born-Infeld term. We show that, by choosing suitable potentials, the energy functional admits a topological lower bound saturated by first-order self-dual equations. The resulting dyonic systems can be divided into vortex-vortex and vortex-antivortex configurations, and the coexistence of vortices and antivortices in the latter extends the vortex-only result known in product Chern-Simons model. On a doubly periodic domain, we establish Bradlow-type bounds with distinct physical implications: for vortex-only systems, the vortex numbers stay below this bound and cannot be arbitrarily large; for vortex-antivortex systems, the bound is imposed on the difference between the vortex and antivortex numbers, while the individual numbers are arbitrary.  This distinction results in a bounded energy spectrum for the former and an unbounded energy spectrum for the latter.
\medskip

\medskip


\medskip


\end{abstract}
\section{Introduction}
\setcounter{equation}{0}

Vortices are represented as soliton-like solutions in the context of the quantum field theory and originate from the study of Type-II superconductors initiated by Abrikosov \cite{AA}. Within the framework of dual string models \cite{HP}, Nielsen and
Olesen established the connection between vortex-line solution of the Abelian Higgs model and high-energy physics. Vortices exert profound implication in condensed-matter and particle physics and associate with many important theoretical issues. For vortices of local topological charges with the same sign, all magnetic fluxlines point in the same direction. However, Yang \cite{Y2} demonstrated that vortices and antivortices can coexist with opposite winding directions in an Abelian gauge model. The study of vortex-antivortex pairs is physically significant. An important application is the classical XY model, where the mutual attraction between vortices and antivortices is described by a logarithmic potential which leads to bind pairs at low temperatures. As the system temperature rises to a critical value, the vortex-antivortex pairs unbind and the system enters a disorder phase. This behavior gives rise to the Berezinskii-Kosterlitz-Thouless phase transition \cite{XY2}. Despite their importance, solving the equations of motion for vortex-only or vortex-antivortex systems, even in the simplest Abelian Higgs model, is difficult due to complexity of field equations. However, it came as a fortune after the famous work of Bogomol'nyi \cite{Bo} and Prasad-Sommerfield \cite{PS}, who developed a BPS structure such that the complicated second-order equations can be simplified into first-order self-dual equations.
The Bogomol'nyi reduction is needed in many areas, such as electroweak theory \cite{AO1,AO2,AO3,AO4,SYew1,SYew2,BL1,BL2,Yew}, Chern--Simons models \cite{HKP,JW,Caff-Y,Han-L-Y,Han-Y2}, cosmic strings \cite{V,CG,Ycosmic1,Ycosmic2,Ybook} and Born-Infeld theories \cite{BI1,BI2,SH,Ybi,Han,XY2}.

The static solutions of the Chern-Simons model are called dyons, which carry electric and magnetic charges. The study of Chern-Simons vortex models is relevant to many areas, including anyons and fractional statistics \cite{FW,DJFA}, fractional quantum Hall effect \cite{DHA} and anyon superconductivity \cite{YFEB}. In \cite{Schroers}, Schroers carries out a detailed derivation of the Bogomol'nyi equations of the general $m$-fold product $U(1)$ Chern-Simons-Higgs theory and considers a special $U(2)$ model. In \cite{HY0}, Han and Yang extend the product Abelian Higgs model of Tong-Wong \cite{DK} and derive a general $U(1) \times U(1)$ Chern-Simons-Higgs model with two Higgs fields $\phi$ and $\psi$ carrying charges $(a,b)$ and $(c,d)$ for any real parameters $a,b,c,d$. They show that the topological energy lower bound is attained when the Bogomol'nyi equations hold and establish an existence result for the multivortex solutions of the equations in the full plane and a doubly-periodic domain, respectively, by following the method in \cite{Y1,XCY}.

The goal of this work is to extend the model in \cite{HY0}, under the same non-degeneracy condition $ad-bc\neq0$, and to develop several new mixed $U(1) \times U(1)$ Chern-Simons models. Following the spirit of \cite{XY2}, which developed product hybrid Born-Infeld models governed by Born-Infeld and Maxwell theories, we first construct the charged vortex-vortex hybrid systems, where one of the $U(1)$ gauge fields is governed by a Chern-Simons term and the other by either a Maxwell term or a Born-Infeld term. We then generalize the product system to accommodate vortex-antivortex pairs by considering distinct combinations: one between the Chern-Simons and Harmonic map models, and another between the Chern-Simons and Born-Infeld models (supporting vortex-antivortex). We show that, when the Higgs potential takes a specific coupled form --- namely, the Chern-Simons potential coupled with the Abelian Higgs, Born-Infeld, or harmonic map potential --- all these mixed models admit a Bogomol'nyi reduction, which leads to energy lower bounds saturated by self-dual equations. In addition, for the resulting systems, we derive exact quantized expressions for the magnetic fluxes, electric charges, and energies. In particular, on the doubly-periodic domain, we obtain Bradlow bounds that impose geometric constraints on the vortex and antivortex numbers or topological charges.

The mixed models constructed in this paper can be understood from two complementary perspectives. One perspective is that, by introducing a Chern-Simons term, our models extend product Abelian Higgs, Born-Infeld, or harmonic map models so that dyons appear. The other is that we extend product Chern-Simons models by replacing a Chern-Simons term with a Maxwell or Born-Infeld term, which allows the coexistence of vortices and antivortices. As a result, the constructed systems give rise to new forms of BPS equations and bring new challenges and opportunities for mathematical analysis.

An outline of the rest of the paper is as follows. In Section \ref{sec2}, we recall some results from \cite{HY0}, which lay the foundation for our present work. In Sections \ref{sec3} and \ref{sec4}, we construct coupled $U(1) \times U(1)$ systems which consist of Maxwell and Chern-Simons vortices, Born-Infeld and Chern-Simons vortices, respectively. In Sections \ref{sec5} and \ref{sec6}, we develop vortex-antivortex models: the Chern-Simons vortices and Maxwell vortices-antivortices, Chern-Simons vortices and Born-Infeld vortices-antivortices, respectively. In Section \ref{sec7}, we obtain the Bradlow bounds of these models on a compact doubly-periodic domain. In Section \ref{sec8}, we draw conclusions.

\section{Bogomol'nyi equations of Chern-Simons models}\lb{sec2}
In this section, we review two models which prepare the ground for our subsequent study: the single-species Chern-Simons theory \cite{HKP,JW} and the product Chern-Simons model developed by Han and Yang \cite{HY0}.
The Lagrangian density of the former defined over the standard Minkowski spacetime $\mathbb{R}^{2,1}$ with the metric $g_{\mu \nu}=\mbox{diag}\{1,-1,-1\}$ is given by
\be\lb{2.1}
{\cal L}=-\frac{\ka}{4} \epsilon ^{\mu \nu \alpha}A_{\mu} F_{\nu \alpha}+D_{\mu}\phi\overline{D^{\mu}\phi}-\frac{1}{\ka^2}|\phi|^2(|\phi|^2-\zeta)^2,
\ee
where $\ka \in \bfR$ is a constant referred to as the Chern-Simons coupling parameter, $\epsilon ^{\mu \nu \alpha}$ is the skew-symmetric Kronecker symbol with $\epsilon ^{012}=1$, $\phi$ is a complex-valued Higgs scalar field, the electromagnetic tensor field $F_{\nu \alpha}=\pa_{\nu}A_{\alpha}-\pa_{\alpha}A_{\nu}$ induced from the real-valued gauge field $A_{\mu}(\mu=0,1,2)$, and the $D_{\mu} \phi =\pa_{\mu} \phi -\ii A_{\mu} \phi$ is gauge-covariant derivative.

The extremals of the Lagrangian density \eq{2.1} satisfy its Euler-Lagrange equations
\bea
\frac{\ka}{2}\epsilon^{\mu \nu \alpha}F_{\nu \alpha}&=&-J^{\mu},\lb{2.2}\\
D_{\mu}D^{\mu} \phi &=&-\frac{1}{\ka^2}(2|\phi|^2(|\phi|^2-\zeta)+(|\phi|^2-\zeta)^2)\phi,\lb{2.3}
\eea
where the current density is given by
\be\lb{2.4}
J^{\mu}=\ri (\phi \overline{D^{\mu}\phi}-\overline{\phi}D^{\mu}\phi),\quad \mu=0,1,2.
\ee
In this paper, we interested in static configurations such that the fields depend only on the planar coordinates $x_1,x_2$ and are independent of the time coordinate, $t=x^{0}$. In the static case, the $\mu=0$ component of \eq{2.4} reduces to the Chern-Simons Gauss law
\be\lb{2.5}
\ka F_{12}=J^{0}=\rho =2A_0|\phi|^2,
\ee
which links the magnetic field $F_{12}$ with the electric charge density $\rho$. This relation leads to the energy
\bea
\mathcal{H}&=&\frac{\ka^2}{4|\phi|^2}F_{12}^2+|D_j\phi|^2+\frac{1}{\ka^2}|\phi|^2(|\phi|^2-\zeta)^2\\
&=&\left(\frac{\ka}{2}\frac{F_{12}}{|\phi|}\pm\frac{|\phi|}{\ka}(|\phi|^2-\zeta)\right)^2\mp F_{12}(|\phi|^2-\zeta)+|D_j\phi|^2.\lb{a2.5}
\eea
Using the relation
\be
|D_i \phi|^2=|D_1 \phi \pm \ri D_2 \phi|^2\pm \ri \left(\pa_1[\phi\overline{D_2 \phi}]-\pa_2[\phi\overline{D_1\phi}]\right)\pm F_{12}|\phi|^2 \lb{3.12}
\ee
in \eq{a2.5}, we obtain
\bea\lb{2.5a}
E&=&\int_{\bfR^2}{\cal H}\,\dd x\nn\\
&=&\int_{\bfR^2}\left(\frac{\ka}{2}\frac{F_{12}}{|\phi|}\pm\frac{|\phi|}{\ka}(|\phi|^2-\zeta)^2\right)\,\dd x+\int_{\bfR^2}|D_1\phi\pm \ri D_2 \phi|^2\,\dd x \pm \int_{\bfR^2}\zeta F_{12}\,\dd x, \nn \\
&\ge& 2\pi\zeta|M|,
\eea
where $M$ is the winding number of the complex scalar field $\phi$ and satisfies $M=\pm |M|$. The lower bound in \eq{2.5a} is saturated by the solutions to the following Bogomol'nyi equations
\bea
D_1\phi\pm \ri D_2 \phi&=&0,\lb{2.6} \\
F_{12}\pm \frac{2}{\ka^2}|\phi|^2(|\phi|^2-\zeta)&=&0.\lb{2.7}
\eea
It may be checked that the solutions of \eq{2.6}-\eq{2.7} satisfy \eq{2.2}-\eq{2.3}. On the other hand, \eq{2.6} and the $\bar{\pa}$-Poincar$\mathrm{\acute{e}}$ \cite{JT} implies that the zeros of $\phi$ are all isolated and of integer multiplicities,
 say $q_1,\dots, q_M$, listed repeatedly to count for multiplicities. Then the substitution $u=\ln|\phi|^2$ and \eq{2.7} transform the system \eq{2.6}-\eq{2.7} into the nonlinear elliptic equation
\be\lb{2.8}
\Delta u=\frac{4}{\ka^2}\e^u(\e^u-1)+4\pi\sum_{s=1}^M \delta_{q_s}(x),\quad x\in\bfR^2,
\ee
where $\delta_q(x)$ is defined over the full plane $\bfR^2$. Yang \cite{Ybook} solved the equation \eq{2.8} in a doubly periodic domain and the full plane, respectively.

In view of \eq{2.5}, \eq{2.7} and \eq{2.8}, we can obtain the expressions of the magnetic flux and electric charge for the $M$-vortex solution:
\be
\Phi=\int_{\bfR^2}F_{12}\,\dd x=2\pi M,\quad Q=\int_{\bfR^2}\rho\,\dd x=2\pi \ka M.
\ee

In \cite{HY0}, Han and Yang extend the Chern-Simons theory \eq{2.1} and construct a product model over the Abelian gauge group $U(1)\times U(1)$, which includes two complex scalar fields $\phi$ and $\psi$, carrying general doublet charges $(a,b)$ and $(c,d)$ for arbitrary $a,b,c,d \in \bfR$, respectively. The Lagrangian density reads
\be\lb{2.9}
\mathcal{L}=-\frac{\ka}{4}\epsilon^{\mu \nu \alpha}\hat{A}_{\mu}\hat{F}_{\nu \alpha}-\frac{\tilde{\ka}}{4}\epsilon^{\mu \nu \alpha}\tilde{A}_{\mu}\tilde{F}_{\nu \alpha}+D_{\mu}\phi\overline{D^{\mu}\phi}+D_{\mu}\psi\overline{D^{\mu}\psi}-V,
\ee
where the potential density is chosen as
\bea
V(|\phi|^2,|\psi|^2)&=&|\phi|^2\left(\left[\frac{a^2}{\ka}+\frac{b^2}{\tilde{\ka}}\right][|\phi|^2-\xi]+\left[\frac{ac}{\ka}+\frac{bd}{\tilde{\ka}}\right][|\psi|^2
-\zeta]\right)^2\nn \\
&&+|\psi|^2\left(\left[\frac{ac}{\ka}+\frac{bd}{\tilde{\ka}}\right][|\phi|^2-\xi]+\left[\frac{c^2}{\ka}+\frac{d^2}{\tilde{\ka}}\right][|\psi|^2-\zeta]\right)^2,
\eea
$\tilde{\ka}>0$ is a coupling constant, and $\xi, \zeta$ are positive constants. The static governing equations are subject to a BPS reduction \cite{HY0}:
\bea
\hat{F}_{12}\pm \frac{2a}{\ka}|\phi|^2\left(\left[\frac{a^2}{\ka}+\frac{b^2}{\tilde{\ka}}\right][|\phi|^2-\xi]+\left[\frac{ac}{\ka}+\frac{bd}{\tilde{\ka}}\right][|\psi|^2
-\zeta]\right),\nn \\
\pm\frac{2c}{\ka}|\psi|^2\left(\left[\frac{ac}{\ka}+\frac{bd}{\tilde{\ka}}\right][|\phi|^2-\xi]+\left[\frac{c^2}{\ka}+\frac{d^2}{\tilde{\ka}}\right][|\psi|^2-\zeta]\right)&=&0, \lb{2.10}\\
\tilde{F}_{12}\pm \frac{2b}{\tilde{\ka}}|\phi|^2\left(\left[\frac{a^2}{\ka}+\frac{b^2}{\tilde{\ka}}\right][|\phi|^2-\xi]+\left[\frac{ac}{\ka}+\frac{bd}{\tilde{\ka}}\right][|\psi|^2
-\zeta]\right)\nn \\
\pm \frac{2d}{\tilde{\ka}}|\psi|^2\left(\left[\frac{ac}{\ka}+\frac{bd}{\tilde{\ka}}\right][|\phi|^2-\xi]+\left[\frac{c^2}{\ka}+\frac{d^2}{\tilde{\ka}}\right][|\psi|^2-\zeta]\right)&=&0,\\ \lb{2.11}
D_1\phi\pm \ri D_2 \phi &=& 0, \\ \lb{2.12}
D_1\psi\pm \ri D_2 \psi &=& 0.\lb{2.13}
\eea
The zeros of the complex scalar fields $\phi$ and $\psi$ are represented by
\be
\phi: q_1,\cdots,q_{M_1}, \quad \psi: \tilde{q}_1,\cdots,\tilde{q}_{M_2},
\ee
respectively. The solutions of the self-dual equations \eq{2.10}-\eq{2.13} saturate the energy lower bound, yielding the energy formula:
\be\lb{2.14}
E=2\pi(\xi M_1+\zeta M_2).
\ee
In the following sections, we shall modify the product Chern-Simons model \eq{2.9} and construct several new charged vortex and vortex-antivortex systems generated from various Bogomol'nyi equations.

\section{System of vortices governed by product  Maxwell and \\Chern--Simons  electrodynamics}\lb{sec3}
Motivated by the mixed Born-Infeld models of \cite{XY2}, this section aims to generalize the product Chern-Simons model \eq{2.9} and develop a hybrid model in which one gauge sector is dominated by the classical Maxwell action while the other maintains the Chern-Simons term. We also show that such models preserve a Bogomol'nyi structure under a suitable choice of the potential.

Let $\hat{A}_i$ and $\tilde{A}_i$ be two real-valued Abelian gauge fields, whose magnetic strengths are $\hat{F}_{ij}=\pa_{i}\hat{A}_j-\pa_j\hat{A}_i$ and $\tilde{F}_{ij}=\pa_i\tilde{A}_j-\pa_j\tilde{A}_i$, respectively. For the complex scalar Higgs fields $\phi$ and $\psi$, carrying charge pairs $(a,b)$ and $(c,d)$, respectively, we defined the gauge-covariant derivatives as
\be\lb{3.0}
D_i \phi =\pa_i \phi -\ri(a\hat{A}_i+b\tilde{A}_i)\phi, \quad \quad D_i \psi =\pa_i \psi -\ri(c\hat{A}_i+d\tilde{A}_i)\psi,\quad i=1,2,
\ee
with the nondegeneracy condition
\be\lb{3.0a}
ad-bc\neq0
\ee
imposed throughout this work.

Following the method developed in \cite{XY2}, the Lagrangian density of the hybrid Chern-Simons model is given by
\be\lb{3.1}
\mathcal{L}=-\frac{1}{4 \lm}\hat{F}_{\mu \nu}\hat{F}^{\mu \nu}-\frac{\ka}{4}\epsilon^{\mu \nu \alpha}\tilde{A}_{\mu}\tilde{F}_{\nu \alpha}+D_{\mu}\phi\overline{D^{\mu}\phi}+D_{\mu}\psi\overline{D^{\mu}\psi}-V,
\ee
where $\lm, \ka>0$ are coupling constants and $V=V(|\phi|^2,|\psi|^2)$ denotes the potential energy density, whose specific form will be fixed by the requirement of admitting a Bogomol'nyi structure. We note that if the Chern-Simons term in \eq{3.1} is replaced by a Maxwell term, the model would reduce to the product Abelian Higgs theory \cite{HY0,Schroers}, whose static solutions are called magnetic vortices and carry no electric charge. In our hybrid model \eq{3.1}, however, we can consider dyon solutions thanks to the presence of the Chern-Simons term. The Euler-Lagrange equations of \eq{3.1} are
\bea
D_{\mu}D^{\mu}\phi &=&-\frac{\pa V}{\pa \overline{\phi}},\lb{3.2}\\
D_{\mu}D^{\mu}\psi &=& -\frac{\pa V}{\pa \overline{\psi}},\lb{3.3}\\
\frac{1}{\lm}\pa_{\nu}\hat{F}^{\mu \nu}&=&-\ri \left(a[\phi\overline{D^{\mu}\phi}-\overline{\phi}D^{\mu}\phi]+c[\psi\overline{D^{\mu}\psi}-\overline{\psi}D^{\mu}\psi]\right),\lb{3.4}\\
\frac{\ka}{2}\epsilon^{\mu \nu \alpha}\tilde{F}_{\nu \alpha}&=&-\ri \left(b[\phi\overline{D^{\mu}\phi}-\overline{\phi}D^{\mu}\phi]+d[\psi\overline{D^{\mu}\psi}-\overline{\psi}D^{\mu}\psi]\right).\lb{3.5}
\eea
The static version of \eq{3.2}-\eq{3.5} read
\bea
D^2_i\phi&=&\frac{\pa V(|\phi|^2,|\psi|^2)}{\pa{\overline{\phi}}}-b^2\tilde{A}_0^2\phi,\lb{3.8}  \\
D^2_i\psi&=&\frac{\pa V(|\phi|^2,|\psi|^2)}{\pa{\overline{\psi}}}-d^2\tilde{A}_0^2\psi, \lb{3.9} \\
\frac{1}{\lm}\pa_j \hat{F}_{ij}&=&\ri(a[\phi\overline{D_i\phi}-\overline{\phi}D_i\phi])+ c[\psi\overline{D_i\psi}-\overline{\psi}D_i\psi]),\lb{3.10}\\
\ka\vep_{ij}\pa_j\tilde{A}_0&=&\ri(b[\phi\overline{D_i\phi}-\overline{\phi}D_i\phi])+ d[\psi\overline{D_i\psi}-\overline{\psi}D_i\psi]),\lb{3.11}\\
\ka\tilde{F}_{12}&=&2(b^2|\phi|^2+d^2|\psi|^2)\tilde{A_{0}}.\lb{3.6}
\eea
The equation \eq{3.6} is the celebrated Gauss law, obtained by setting the component $\mu=0$ in \eq{3.5}. In contrast to \eq{2.5},  \eq{3.6} involves two charged scalar fields $\phi$ and $\psi$.
We know that the topological invariant generated by the Chern-Simons term is independent of the spacetime metric $g_{\mu\nu}$. This fact implies that this term does not arise in the energy-momentum tensor $T_{\mu \nu}$ of the action density \eq{3.1}, which leads to the tensor
\be
T_{\mu \nu}=-\frac{1}{\lm}g^{\beta \beta'}\hat{F}_{\mu \beta}\hat{F}_{\nu \beta'}+D_{\mu}\phi\overline{D_{\nu}\phi}+\overline{D_{\mu }\phi}D_{\nu}\phi+ D_{\mu} \psi\overline{D_{\nu}\psi}+\overline{D_{\mu}\psi}D_{\nu}\psi-g_{\mu \nu}{\cal L}_0,
\ee
where ${\cal L}_0$ is defined by taking $\ka=0$ in the Lagrangian \eq{3.1}. Thus we obtain that, in the static situation where one sector maintains electrically neutral while the other is charged, say $\hat{A}_{0}=0$ and $\tilde{A_0}\neq 0$, the planar-space Hamiltonian energy density is given by
\bea\lb{3.7}
\mathcal{H}&=&T_{00}\nn \\
&=&\frac{1}{2 \lm}\hat{F}_{12}^2+(b^2|\phi|^2+d^2|\psi|^2)\tilde{A}_0^2+|D_i\phi|^2+|D_i\psi|^2+V.
\eea
On the other hand, it is worth noting that the following identities hold:
\bea
|D_i \phi|^2=|D_1 \phi \pm \ri D_2 \phi|^2\pm \ri \left(\pa_1[\phi\overline{D_2 \phi}]-\pa_2[\phi\overline{D_1\phi}]\right)\pm (a\hat{F}_{12}+b\tilde{F}_{12})|\phi|^2,\lb{3.13a}\\
|D_i \psi|^2=|D_1 \psi \pm \ri D_2 \psi|^2\pm \ri \left(\pa_1[\psi\overline{D_2 \psi}]-\pa_2[\psi\overline{D_1\psi}]\right)\pm (c\hat{F}_{12}+d\tilde{F}_{12})|\psi|^2,\lb{3.13}
\eea
which are similar to the relation \eq{3.12}.
As a result, combining \eq{3.6}, \eq{3.13a} and \eq{3.13}, the energy density may be rewritten as
\bea\lb{3.14}
\mathcal{H}
&=&\frac{1}{2\lm}\hat{F}_{12}^2+\frac{\ka^2\tilde{F}_{12}^2}{4(b^2|\phi|^2+d^2|\psi|^2)}+|D_i\phi|^2+|D_i \psi|^2+V \nn \\
&=&\frac{1}{2\lm}\left(\hat{F}_{12}\pm \lm (a(|\phi|^2-\xi)+c(|\psi|^2-\zeta)\right)^2\mp\hat{F}_{12}(a(|\phi|^2-\xi)+c(|\psi|^2-\zeta))\nn \\
&&+\left(\frac{\ka \tilde{F}_{12}}{2W}\pm \frac{1}{\ka}W(b(|\phi|^2-\xi)+d(|\psi|^2-\zeta))\right)^2\mp\tilde{F}_{12}(b(|\phi|^2-\xi)+d(|\psi|^2-\zeta))\nn \\
&&+|D_1 \phi \pm \ri D_2 \phi|^2 \pm \ri \left(\pa_1[\phi\overline{D_2 \phi}]-\pa_2[\phi\overline{D_1\phi}]\right)\pm (a\hat{F}_{12}+b\tilde{F}_{12})|\phi|^2 \nn \\
&&+|D_1 \psi \pm \ri D_2 \psi|^2 \pm \ri \left(\pa_1[\psi\overline{D_2 \psi}]-\pa_2[\psi\overline{D_1\psi}]\right)\pm (c\hat{F}_{12}+d\tilde{F}_{12})|\psi|^2\nn \\
&&-\frac{\lm}{2}\left(a(|\phi|^2-\xi)+c(|\psi|^2-\zeta)\right)^2-\frac{1}{\ka^2}W^2\left(b(|\phi|^2-\xi)+d(|\psi|^2-\zeta)\right)^2+V,
\eea
where
\be\lb{3.14a}
W=\sqrt{b^2|\phi|^2+d^2|\psi|^2},
\ee
$\xi, \zeta>0$ are constants and $\sqrt{\xi},\sqrt{\zeta}$ describe the energy scales of the spontaneously broken symmetries. To reveal a Bogomol'nyi structure, we choose the potential density
\bea
V(|\phi|^2,|\psi|^2)&=&\frac{\lm}{2}\left(a(|\phi|^2-\xi)+c(|\psi|^2-\zeta)\right)^2 \nn \\
&&+\frac{b^2|\phi|^2+d^2|\psi|^2}{\ka^2}\left(b(|\phi|^2-\xi)+d(|\psi|^2-\zeta)\right)^2 \lb{3.14b}
\eea
We observe that the first term on the right-hand side of \eq{3.14b} is the potential density of classical Abelian Higgs type, and the second term is of the Chern-Simons type. Inserting \eq{3.14b} into \eq{3.14}, we get
\bea
\mathcal{H}
&=&\frac{1}{2\lm}\left(\hat{F}_{12}\pm \lm (a(|\phi|^2-\xi)+c(|\psi|^2-\zeta)\right)^2+|D_1 \phi \pm \ri D_2 \phi|^2  \nn \\
&&+\left(\frac{\ka \tilde{F}_{12}}{2W}\pm \frac{1}{\ka}W(b(|\phi|^2-\xi)+d(|\psi|^2-\zeta))\right)^2+|D_1 \psi \pm \ri D_2 \psi|^2\nn \\
&&\pm \ri \left(\pa_1[\phi\overline{D_2 \phi}]-\pa_2[\phi\overline{D_1\phi}]\right)\pm \ri \left(\pa_1[\psi\overline{D_2 \psi}]-\pa_2[\psi\overline{D_1\psi}]\right)  \nn \\
&&\pm (a\xi+c\zeta)\hat{F}_{12}\pm (b\xi+d\zeta)\tilde{F}_{12},
\eea
which results in
\be\lb{3.15}
E=\int_{\bfR^2} \mathcal{H}\,\dd x\ge\pm(a\xi+c\zeta) \int_{\bfR^2}\hat{F}_{12}\,\dd x\pm \int_{\bfR^2}(b\xi+d\zeta)\tilde{F}_{12}\,\dd x,
\ee
which renders us to derive the first-order self-dual (anti-self-dual) equations
\bea
\hat{F}_{12}\pm \lm \left(a(|\phi|^2-\xi)+c(|\psi|^2-\zeta)\right)&=&0,\lb{3.16}\\
\tilde{F}_{12}\pm \frac{2}{\ka^2}(b^2|\phi|^2+d^2|\psi|^2)\left(b(|\phi|^2-\xi)+d(|\psi|^2-\zeta)\right)&=&0,\lb{3.17}\\
D_1\phi\pm \ri D_2 \phi &=& 0, \lb {3.18}\\
D_1\psi\pm \ri D_2 \psi &=& 0.\lb{3.19}
\eea
A simple calculation of \eq{3.16} and \eq{3.17} gives rise to the expressions:
\bea
\hat{F}_{12}&=&\mp\lm(a(|\phi|^2-\xi)+c(|\psi|^2-\zeta)),\lb{3.20}\\
\tilde{F}_{12}&=&\mp \frac{2}{\ka^2}(b^2|\phi|^2+d^2|\psi|^2)\left(b(|\phi|^2-\xi)+d(|\psi|^2-\zeta)\right),\lb{3.21}
\eea
respectively. Besides, it is straightforward to check that the solutions of \eq{3.16}-\eq{3.19} satisfy \eq{3.8}-\eq{3.6}.

Using
\be\lb{3.22}
\phi: \left\{q'_1,\dots,q'_{M_1}\right\},\quad \psi:\left\{{q}''_1,\dots,{q}''_{M_2}\right\},
\ee
to denote the zeros of the complex scalar fields $\phi$ and $\psi$, respectively, with repetitions counting for  multiplicities. Combining the equations \eq{3.18} and \eq{3.19}, we arrive at
\be\lb{3.23}
a\hat{F}_{12}+b\tilde{F}_{12}=\mp \frac12 \Delta \ln |\phi|^2, \quad c\hat{F}_{12}+d\tilde{F}_{12}=\mp \frac12 \Delta \ln |\psi|^2,
\ee
away from the zeros of $\phi$ and $\psi$.
Then, using the substitutions $u=\ln|\phi|^2$, $v=\ln|\psi|^2$, along with \eq{3.20}, \eq{3.21} and \eq{3.23}, we obtain the coupled nonlinear elliptic equations
\bea
\Delta u&=&2\lm(a^2(\e^u-\xi)+ac(\e^v-\zeta))\nn \\
&&+\frac{4b}{\ka^2}(b^2\e^u+d^2\e^v)(b(\e^u-\xi)+d(\e^v-\zeta))+4\pi\sum_{s=1}^{M_1} \delta_{q'_s},\lb{3.24}\\
\Delta v&=&2\lm (ac(\e^u-\xi)+c^2(\e^v-\zeta))\nn \\
&&+\frac{4d}{\ka^2}(b^2\e^u+d^2\e^v)(b(\e^u-\xi)+d(\e^v-\zeta))+4\pi\sum_{s=1}^{M_2} \delta_{q''_s}.\lb{3.25}
\eea

We next calculate the magnetic fluxes and energy of this vortex system, where the vortices are characterized by the zero sets of the Higgs fields $\phi$ and $\psi$ given in \eq{3.22}.
To avoid inconvenience generated from the boundary terms, we here study \eq{3.24} and \eq{3.25} in a doubly-periodic domain, $\Omega$, which satisfies the t' Hooft boundary conditions \cite{WY,Hooft} and is of independent interest as it depicts Abrikosov's vortex condensates \cite{Ab}.
To this end, we introduce the background functions $u_0^1$ and $u_0^2$ (cf. \cite{Aubin}) satisfying
\be\lb{3.26}
\Delta u_0^1=-\frac{4 \pi M_1}{|\Om|}+4 \pi \sum_{s=1}^{M_1} \delta_{q'_{s}},\quad \Delta u_0^2=-\frac{4 \pi M_2}{|\Om|}+4 \pi \sum_{s=1}^{M_2} \delta_{{q}''_{s}}.
\ee
Inserting the variable transformation $u=u_0^1+w_1$ and $v=u_0^2+w_2$ into \eq{3.24} and \eq{3.25}, we arrive at the following source-free equations
\bea
\Delta w_1&=&2\lm(a^2(\e^{u_0^1+w_1}-\xi)+ac(\e^{u_0^2+w_2}-\zeta))\nn \\
&&+\frac{4b}{\ka^2}(b^2\e^{u_0^1+w_1}+d^2\e^{u_0^2+w_2})\left(b(\e^{u_0^1+w_1}-\xi)+d(\e^{u_0^2+w_2}-\zeta)\right)
+\frac{4 \pi M_1}{|\Om|},\nn \\
\lb{3.27}\\
\Delta w_2&=&2\lm (ac(\e^{u_0^1+w_1}-\xi)+c^2(\e^{u_0^2+w_2}-\zeta))\nn \\
&&+\frac{4d}{\ka^2}(b^2\e^{u_0^1+w_1}+d^2\e^{u_0^2+w_2})\left(b(\e^{u_0^1+w_1}-\xi)+d(\e^{u_0^2+w_2}-\zeta)\right)+\frac{4 \pi M_2}{|\Om|}.\nn \\
\lb{3.28}
\eea
Integrating \eq{3.27} and \eq{3.28} gives the quantized integrals
\bea
\lm\int f_1\, \dd x&=&\frac{2\pi (b M_2-d M_1)}{ad-bc},\lb{3.29}\\
\int\frac{2}{\ka^2}(b^2\e^{u_0^1+w_1}+d^2\e^{u_0^2+w_2})f_2\, \dd x&=&\frac{2\pi (c M_1-a M_2)}{ad-bc}, \lb{3.30}
\eea
where the quantities $f_1$ and $f_2$ are defined by
\bea
{f}_1&=&f_1(u_0^1+w_1,u_0^2+w_2)=a(\e^{u_0^1+w_1}-\xi)+c(\e^{u_0^2+w_2}-\zeta),\lb{3.31}\\
{f}_2&=&f_2(u_0^1+w_1,u_0^2+w_2)=b(\e^{u_0^1+w_1}-\xi)+d(\e^{u_0^2+w_2}-\zeta).\lb{3.32}
\eea
Substituting \eq{3.29} and \eq{3.30} into \eq{3.20} and \eq{3.21}, we obtain the fluxes
\bea
\hat{\Phi}&=&\int\hat{F}_{12}\, \dd x=\pm \frac{2\pi(d M_1-b M_2)}{ad-bc},\lb{3.33} \\
\tilde{\Phi}&=&\int\tilde{F}_{12}\, \dd x=\pm \frac{2\pi(a M_2-c M_1)}{ad-bc}.\lb{3.34}
\eea
Combining \eq{3.6}, \eq{3.15}, \eq{3.33}, and \eq{3.34}, we arrive at the expressions of the energy and electric charge as follows:
\bea
E&=&\int\mathcal{H}\, \dd x=2\pi \left(\xi M_1+\zeta M_2\right), \lb{3.35} \\
Q&=&\int \ka \tilde{F}_{12}\, \dd x= \pm \frac{2\pi\ka(a M_2-c M_1)}{ad-bc}. \lb{3.36}
\eea
From \eq{3.35} and \eq{3.36}, we observe that the energy depends only on the vortex numbers $M_1$ and $M_2$, not on the domain $\Om$ chosen or the charge parameters $a,b,c,d$. In contrast, the electric charge is jointly determined by the vortex numbers and charge parameters.

Recall that in the static product Abelian Higgs theory, only magnetic vortices appear while dyons are absent. The Chern-Simons term is what enables dyon solutions, and this key feature is preserved in our hybrid construction \eq{3.7} equipped with the potential \eq{3.14b}. The derived equations \eq{3.16}-\eq{3.19} reveal how the Chern-Simons charged vortices and Maxwell vortices interact with each other, and these equations pave the way for vortex-antivortex investigations presented later.

\section{System of vortices governed by two-species Chern-Simons and Born-Infeld electrodynamics}\lb{sec4}
This section is devoted to obtaining Bogomol'nyi equations for a system of vortices governed by product Chern-Simons and Born-Infeld theories. The static energy density in this case is given by
\be\lb{4.7}
\mathcal{H}=b_1^2\left(\sqrt{1+\frac{1}{b_1^2}\hat{F}_{12}^2}-1\right)+(b^2|\phi|^2+d^2|\psi|^2)\tilde{A}_0^2+|D_i\phi|^2+|D_i\psi|^2+V,
\ee
where $b_1>0$ is Born-Infeld coupling constant. We see from \eq{4.7} that this model becomes the product Born-Infeld theory constructed by Xu and Yang \cite{XY2} after replacing the second term with Born-Infeld nonlinear electrodynamics. Also, this highly nonlinear mixed model also allows us to consider dyons. Moreover, comparing \eq{3.7} and \eq{4.7}, we see that the Maxwell term in \eq{3.7} has been replaced by the Born-Infeld term. The Euler-Lagrange equations of \eq{4.7} are
\bea
D^2_i\phi&=&\frac{\pa V(|\phi|^2,|\psi|^2)}{\pa{\overline{\phi}}}-b^2\tilde{A}_0^2\phi,\lb{a4.8}  \\
D^2_i\psi&=&\frac{\pa V(|\phi|^2,|\psi|^2)}{\pa{\overline{\psi}}}-d^2\tilde{A}_0^2\psi, \lb{a4.9} \\
\pa_j \left(\frac{\hat{F}_{ij}}{\sqrt{1+\frac{1}{b_1^2}\hat{F}_{12}^2}}\right)&=&\ri(a[\phi\overline{D_i\phi}-\overline{\phi}D_i\phi])+ c[\psi\overline{D_i\psi}-\overline{\psi}D_i\psi]),\lb{a4.10}\\
\ka\vep_{ij}\pa_j\tilde{A}_0&=&\ri(b[\phi\overline{D_i\phi}-\overline{\phi}D_i\phi])+ d[\psi\overline{D_i\psi}-\overline{\psi}D_i\psi]),\lb{a4.11}\\
\ka\tilde{F}_{12}&=&2(b^2|\phi|^2\tilde{A_0}+d^2|\psi|^2)\tilde{A_{0}}.\lb{a4.12}
\eea
From \eq{3.6} and \eq{a4.12}, we observe that the models \eq{3.7} and \eq{4.7} satisfy the same Gauss law. This is expected, since the Chern-Simons action remains unchanged.
Substituting \eq{3.13a}, \eq{3.13} and \eq{a4.12} into \eq{4.7} and reorganize the Hamiltonian as
\bea\lb{4.8}
\mathcal{H}&=&b_1^2\left(\sqrt{1+\frac{1}{b_1^2}\hat{F_{12}^2}}-1\right)+\frac{\ka^2\tilde{F}_{12}^2}{4(b^2|\phi|^2+d^2|\psi|^2)}+|D_i\phi|^2+|D_i\psi|^2+V\nn \\
&=&\frac12\left(\hat{F_{12}}\pm F(a(|\phi|^2-\xi)+c(|\psi|^2-\zeta)\right)^2F^{-1}+\frac{b_1^2}{2}(FU-1)^2F^{-1}\nn \\
&&-b_1^2+b_1^2U \mp\hat{F}_{12}\left(a(|\phi|^2-\xi)+c(|\psi|^2-\zeta)\right)\nn \\
&&+\left(\frac{\ka\tilde{F}_{12}}{2W}\pm \frac{W}{\ka}(b(|\phi|^2-\xi)+d(|\psi|^2-\zeta))\right)^2\mp \tilde{F}_{12}\left(b(|\phi|^2-\xi)+d(|\psi|^2-\zeta)\right)\nn \\
&&+|D_1 \phi \pm \ri D_2 \phi|^2\pm \ri \left(\pa_1[\phi\overline{D_2 \phi}]-\pa_2[\phi\overline{D_1\phi}]\right)\pm (a\hat{F}_{12}+b\tilde{F}_{12})|\phi|^2\nn \\
&&+|D_1 \psi \pm \ri D_2 \psi|^2\pm \ri \left(\pa_1[\psi\overline{D_2 \psi}]-\pa_2[\psi\overline{D_1\psi}]\right)\pm (c\hat{F}_{12}+d\tilde{F}_{12})|\psi|^2\nn \\
&&-\frac{W^2}{\ka^{2}}\left(b(|\phi|^2-\xi)+d(|\psi|^2-\zeta)\right)^2+V,
\eea
where $W$ is defined as \eq{3.14a} and
\be
F={\sqrt{1+\frac1{b_1^2}\hat{F}_{12}^2}},\quad U=\sqrt{1-\frac{(a(|\phi|^2-\xi)+c(|\psi|^2-\zeta))^2}{b_1^2}}.\nn
\ee
As a result, if we take the potential density function as
\bea\lb{4.9}
V(|\phi|^2,|\psi|^2)&=& b_1^2(1-U)+\frac{W^2}{\ka^{2}}\left(b(|\phi|^2-\xi)+d(|\psi|^2-\zeta)\right)^2\nn \\
&=&b_1^2\left(1-\sqrt{1-\frac{(a(|\phi|^2-\xi)+c(|\psi|^2-\zeta))^2}{b_1^2}}\right)\nn \\
&&+\frac{(b^2|\phi|^2+d^2|\psi|^2)}{\ka^2}\left(b(|\phi|^2-\xi)+d(|\psi|^2-\zeta)\right)^2,
\eea
which yields the simplified Hamiltonian
\bea\lb{4.10}
\mathcal{H}&=&\frac12\left(\hat{F}_{12}\pm F(a(|\phi|^2-\xi)+c(|\psi|^2-\zeta)\right)^2F^{-1}+\frac{b_1^2}{2}(FU-1)^2F^{-1}\nn \\
&&+\left(\frac{\ka\tilde{F}_{12}}{2W}\pm \frac{W}{\ka}(b(|\phi|^2-\xi)+d(|\psi|^2-\zeta))\right)^2\nn \\
&&+|D_1 \phi \pm \ri D_2 \phi|^2\pm \ri \left(\pa_1[\phi\overline{D_2 \phi}]-\pa_2[\phi\overline{D_1\phi}]\right) \nn \\
&&+|D_1 \psi \pm \ri D_2 \psi|^2\pm \ri \left(\pa_1[\psi\overline{D_2 \psi}]-\pa_2[\psi\overline{D_1\psi}]\right) \nn \\
&&\pm (a\xi+c\zeta)\hat{F}_{12}\pm (b\xi+d\zeta)\tilde{F}_{12}.
\eea
Therefore the energy satisfies the following lower bound over the full domain of interest:
\be\lb{4.11}
E=\int\mathcal{H}\, \dd x\ge\pm (a\xi+c\zeta)\int\hat{F}_{12}\, \dd x\pm(b\xi+d\zeta)\int\tilde{F}_{12}\, \dd x.
\ee
 The lower bound is derived if and only if the field configuration $(\phi,\psi, \hat{A}_j,\tilde{A}_j)$ satisfies the self-dual (anti-self-dual) equations
\bea
\hat{F}_{12}\pm F[a(|\phi|^2-\xi)+c(|\psi|^2-\zeta)]&=&0,\lb{4.12}\\
FU-1&=&0,\lb{4.13}\\
\frac{\ka\tilde{F}_{12}}{2W}\pm\frac{W}{\ka}(b(|\phi|^2-\xi)+d(|\psi|^2-\zeta))&=&0,\lb{4.14}\\
D_1\phi\pm \ri D_2 \phi &=& 0, \lb {4.15}\\
D_1\psi\pm \ri D_2 \psi &=& 0.\lb{4.16}
\eea
Combining \eq{4.12} and \eq{4.13} and rewriting \eq{4.14}, we have the following two vorticity field equations
\bea
\hat{F}_{12}&=&\mp\frac{a(|\phi|^2-\xi)+c(|\psi|^2-\zeta)}{\sqrt{1-\frac{(a(|\phi|^2-\xi)+c(|\psi|^2-\zeta))^2}{b_1^2}}},\lb{4.17}\\
\tilde{F}_{12}&=&\mp \frac{2(b^2|\phi|^2+d^2|\psi|^2)}{\ka^2}(b(|\phi|^2-\xi)+d(|\psi|^2-\zeta)).\lb{4.18}
\eea
As a consequence, in light of the zeros assigned to the complex scalar fields $\phi$ and $\psi$ by \eq{3.22} and the relations in \eq{3.23}, and introducing the substitutions $u=\ln|\phi|^2$ and $v=\ln|\psi|^2$, the BPS equations \eq{4.12}-\eq{4.16} reduce into the source-free equations
\bea
\Delta u&=&2\frac{a^2(\e^u-\xi)+ac(\e^v-\zeta)}{\sqrt{1-\frac{1}{b_1^2}\left(a(\e^u-\xi)+c(\e^v-\zeta)\right)^2}}\nn \\
&&+\frac{4b}{\ka^2}(b^2\e^u+d^2\e^v)(b(\e^u-\xi)+d(\e^v-\zeta))+4\pi\sum_{s=1}^{M_1} \delta_{q'_s},\lb{4.19}\\
\Delta v&=&2\frac{ac(\e^u-\xi)+c^2(\e^v-\zeta)}{\sqrt{1-\frac{1}{b_1^2}\left(a(\e^u-\xi)+c(\e^v-\zeta)\right)^2}}\nn \\
&&+\frac{4d}{\ka^2}(b^2\e^u+d^2\e^v)(b(\e^u-\xi)+d(\e^v-\zeta))+4\pi\sum_{s=1}^{M_2} \delta_{q''_s}.\lb{4.20}
\eea
Following the same procedure as in Section \ref{sec3}, we may obtain from \eq{4.19} and \eq{4.20} the quantized integrals
\bea
a\int\frac{f_1}{\sqrt{1-\frac{1}{b_1^2}f_1^2}}\, \dd x+b\int\frac{2}{\ka^2}(b^2\e^{u}+d^2\e^{v})f_2\, \dd x
&=&-2\pi M_1, \lb{4.23}\\
c\int\frac{f_1}{\sqrt{1-\frac{1}{b_1^2}f_1^2}}\, \dd x+d\int\frac{2}{\ka^2}(b^2\e^{u}+d^2\e^{v})f_2\, \dd x
&=&-2\pi M_2,\lb{4.24}
\eea
where the quantities $f_1$ and $f_2$ satisfy the same expression as \eq{3.31} and \eq{3.32}, respectively.
Upon solving these equations, we rederive the expression \eq{3.30} as well as
\be
\int\frac{f_1}{\sqrt{1-\frac{1}{b_1^2}f_1^2}}\, \dd x=\frac{2\pi (b M_2-d M_1)}{ad-bc}.\lb{4.25}\\
\ee
Substituting \eq{3.30} and \eq{4.25} into \eq{4.17} and \eq{4.18}, we arrive at the magnetic fluxes:
\bea
\hat{\Phi}&=&\int\hat{F}_{12}\, \dd x=\pm \frac{2\pi(d M_1-b M_2)}{ad-bc},\lb{4.27} \\
\tilde{\Phi}&=&\int\tilde{F}_{12}\, \dd x=\pm \frac{2\pi(a M_2-c M_1)}{ad-bc}.\lb{4.28}
\eea
Combining \eq{a4.12}, \eq{4.11}, \eq{4.27}, and \eq{4.28}, we obtain the same energy and electric charge as \eq{3.35} and \eq{3.36}, respectively. Furthermore, we also observe that the models \eq{3.7} and \eq{4.7} have the same magnetic fluxes, as presented in \eq{4.27} and \eq{4.28}.

Like the pure Chern-Simons model \eq{2.9}, our hybrid models \eq{3.7} and \eq{4.7} admit dyon solutions. In fact, by choosing mixed-type potentials \eq{3.14b} and \eq{4.9}, we obtain new Bogomol'nyi equations that present new structures, which motivates one to seek new mathematical techniques to establish existence results. Besides, although the form of the action density is different, these new systems \eq{3.7} and \eq{4.7} have the same expressions for energy and fluxes as those of \eq{2.9}.
\section{Bogomol'nyi equations for Chern-Simons vortices coupled with Maxwell vortices and antivortices}\lb{sec5}
In Sections \ref{sec3} and \ref{sec4}, we obtained the Bogomol'nyi equations of vortex systems governed by product Maxwell-Chern-Simons and Chern-Simons-Born-Infeld theories, respectively. In this section, we turn to Bogomol'nyi equations that accommodate both charged vortices and vortex-antivortex pairs in mixed Chern-Simons and Harmonic map models.

The corresponding Lagrangian density is
\be\lb{5.1}
\mathcal{L}=-\frac{1}{4 \lm}\hat{F}_{\mu \nu}\hat{F}^{\mu \nu}-\frac{\ka}{4}\epsilon^{\mu \nu \alpha}\tilde{A}_{\mu}\tilde{F}_{\nu \alpha}+\frac{2}{(1+|\phi|^2)^2}D_{\mu}\phi\overline{D^{\mu}\phi}+D_{\mu}\psi\overline{D^{\mu}\psi}-V(|\phi|^2,|\psi|^2),
\ee
which leads to the energy-momentum tensor
\be\lb{5.2a}
T_{\mu \nu}=-\frac{1}{\lm}\eta^{\beta \beta'}\hat{F}_{\mu \beta}\hat{F}_{\nu \beta'}+\frac{2}{(1+|\phi|^2)^2}\left(D_{\mu}\phi\overline{D_{\nu}\phi}+\overline{D_{\mu }\phi}D_{\nu}\phi\right)+ D_{\mu} \psi\overline{D_{\nu}\psi}+\overline{D_{\mu}\psi}D_{\nu}\psi-\eta_{\mu \nu}{\cal L}_0,
\ee
where ${\cal L}_0$ follows from the Lagrangian \eq{5.1} by setting $\ka=0$. When we replace the second term in \eq{5.1} with a Maxwell term and let the fourth term take the same form as the third term, the model reduces the product harmonic map model \cite{HHY} that supports vortex-antivortex pairs.
A direct calculation yields the equations of motion
\bea
D_{\mu}\left(\frac{D^{\mu}\phi}{(1+|\phi|^2)^2}\right) &=&-\frac{2D_{\mu}\phi\overline{D^{\mu}\phi}}{(1+|\phi|^2)^3}\phi-\frac12\frac{\pa V}{\pa \overline{\phi}},\lb{5.2}\\
D_{\mu}D^{\mu}\psi &=& -\frac{\pa V}{\pa \overline{\psi}},\lb{5.3}\\
\frac{1}{\lm}\pa_{\nu}\hat{F}^{\mu \nu}&=& -\ri\left( \frac{2a[\phi\overline{D^{\mu}\phi}-\overline{\phi}D^{\mu}\phi]}{(1+|\phi|^2)^2}+ c[\psi\overline{D^{\mu}\psi}-\overline{\psi}D^{\mu}\psi]\right),\lb{5.4}\\
\frac{\ka}{2}\epsilon^{\mu \nu \alpha}\tilde{F}_{\nu \alpha}&=&-\ri\left( \frac{2b[\phi\overline{D^{\mu}\phi}-\overline{\phi}D^{\mu}\phi]}{(1+|\phi|^2)^2} +d[\psi\overline{D^{\mu}\psi}-\overline{\psi}D^{\mu}\psi]\right).\lb{5.5}
\eea
From \eq{5.2a}, we obtain the static Hamiltonian energy density
\bea\lb{5.7}
\mathcal{H}&=&T_{00}\nn \\
&=&\frac{1}{2 \lm}\hat{F}_{12}^2+\frac{2b^2|\phi|^2\tilde{A}_0^2}{(1+|\phi|^2)^2}+d^2|\psi|^2\tilde{A_0^2}+\frac{2|D_i\phi|^2}{(1+|\phi|^2)^2}+|D_i\psi|^2+V,
\eea
whose Euler-Lagrange equations are given by
\bea
D_i\left(\frac{D_i\phi}{(1+|\phi|^2)^2}\right) &=&-\frac{2|D_i\phi|^2}{(1+|\phi|^2)^3}\phi+\frac12\frac{\pa V}{\pa \overline{\phi}}+\frac{2b^2\tilde{A}_0^2|\phi|^2\phi}{(1+|\phi|^2)^3}-\frac{b^2\tilde{A}_0^2\phi}{(1+|\phi|^2)^2},\lb{5.9}\\
D_i^2\psi &=& \frac{\pa V}{\pa \overline{\psi}}-d^2\tilde{A}_0^2\psi,\lb{5.10}\\
\frac{1}{\lm}\pa_j\hat{F}_{ij}&=& \ri\left( \frac{2a[\phi\overline{D_i\phi}-\overline{\phi}D_i\phi]}{(1+|\phi|^2)^2}+ c[\psi\overline{D_i\psi}-\overline{\psi}D_i\psi]\right),\lb{5.11}\\
\ka\vep_{ij}\pa_j\tilde{A}_0&=& \ri\left( \frac{2b[\phi\overline{D_i\phi}-\overline{\phi}D_i\phi]}{(1+|\phi|^2)^2} +d[\psi\overline{D_i\psi}-\overline{\psi}D_i\psi]\right),\lb{5.12}\\
\ka\tilde{F}_{12}&=&2\tilde{A_0}\left(\frac{2b^2|\phi|^2}{(1+|\phi|^2)^2}+d^2|\psi|^2\right),\lb{5.6}
\eea
which are static version of \eq{5.2}-\eq{5.5}. On the other hand, \eq{5.6} is the Gauss law of the model \eq{5.1}, obtained by setting $\mu=0$ in \eq{5.5}.

To uncover a Bogomol'nyi structure, we introduce a current density
\be\lb{5.13}
J_i(\phi)=\frac{\ri}{1+|\phi|^2}(\phi\overline{D_i\phi}-\overline{\phi}D_i\phi),\quad i=1,2,
\ee
which leads to the expression
\be\lb{5.14}
J_{12}(\phi)=\frac{-2|\phi|^2}{1+|\phi|^2}(a \hat{F}_{12}+b\tilde{F}_{12})+\frac{2\ri}{(1+|\phi|^2)^2}(D_1 \phi\overline{D_2 \phi}-\overline{D_1\phi}D_2 \phi).
\ee
On the other hand, note that the identity \eq{3.12} can rewrite as
\be
|D_1\phi|^2+|D_2\phi|^2=|D_1\phi\pm\ii D_2\phi|^2\pm \ri(D_1\phi \overline{D_2\phi}-\overline{D_1\phi}D_2\phi). \lb{5.14a}
\ee
Thus combining \eq{3.13}, \eq{5.6} and \eq{5.14a}, we reorganize the Hamiltonian density \eq{5.7} as
\bea\lb{5.15}
\cal{H}&=&\frac{1}{2 \lm}\hat{F}_{12}^2+\frac{\ka^2\tilde{F}_{12}^2}{4G^2}+\frac{2|D_i\phi|^2}{(1+|\phi|^2)^2}+|D_i\psi|^2+V,\nn \\
&=&\frac{1}{2\lm}\left(\hat{F}_{12}\pm \lm\left(a\frac{|\phi|^2-1}{|\phi|^2+1}+c(|\psi|^2-\zeta)\right)\right)^2\mp\hat{F}_{12}\left(a\frac{|\phi|^2-1}{|\phi|^2+1}+c(|\psi|^2-\zeta)\right)\nn \\
&&+\left(\frac{\ka\tilde{F}_{12}}{2G}\pm\frac{1}{\ka}G\left(b\frac{|\phi|^2-1}{|\phi|^2+1}+d(|\psi|^2-\zeta)\right)\right)^2\mp \tilde{F}_{12} \left(b\frac{|\phi|^2-1}{|\phi|^2+1}+d(|\psi|^2-\zeta)\right)\nn \\
&&+\frac{2}{(1+|\phi|^2)^2}|D_1\phi \pm \ri D_2\phi|^2\pm \frac{2\ri}{(1+|\phi|^2)^2}(D_1\phi\overline{D_2\phi}-\overline{D_1\phi}D_2\phi)\nn \\
&&+|D_1\psi \pm \ri D_2\psi|^2\pm\ri \left(\pa_1[\psi\overline{D_2 \psi}]-\pa_2[\psi\overline{D_1\psi}]\right)\pm(c\hat{F}_{12}+d\tilde{F}_{12})|\psi|^2\nn \\
&&-\frac{\lm}{2}\left(a\frac{|\phi|^2-1}{|\phi|^2+1}+c(|\psi|^2-\zeta)\right)^2-\frac{G^2}{\ka^2}\left(b\frac{|\phi|^2-1}{|\phi|^2+1}+d(|\psi|^2-\zeta)\right)^2
+V,
\eea
where $G$ is defined by
\be\lb{5.8}
G=\sqrt{\frac{2b^2|\phi|^2}{(1+|\phi|^2)^2}+d^2|\psi|^2}.
\ee
In order to derive the Bogomol'nyi equations, we need to choose the potential density $V$ to be
\be\lb{a5.16}
V=\frac{\lm}{2}\left(a\frac{|\phi|^2-1}{|\phi|^2+1}+c(|\psi|^2-\zeta)\right)^2+\frac{G^2}{\ka^2}\left(b\frac{|\phi|^2-1}{|\phi|^2+1}+d(|\psi|^2-\zeta)\right)^2.
\ee
We know from \eq{a5.16} that this potential allows the coexistence of vortices and antivortices --- a feature that is absent in the pure Chern-Simons model \eq{2.9}.
Inserting \eq{5.14} and \eq{a5.16} into \eq{5.15}, we have
\bea\lb{5.16}
\cal{H}&=&\frac{1}{2\lm}\left(\hat{F}_{12}\pm \lm\left(a\frac{|\phi|^2-1}{|\phi|^2+1}+c(|\psi|^2-\zeta)\right)\right)^2\nn \\
&&+\left(\frac{\ka\tilde{F}_{12}}{2G}\pm\frac{1}{\ka}G\left(b\frac{|\phi|^2-1}{|\phi|^2+1}+d(|\psi|^2-\zeta)\right)\right)^2+\frac{2}{(1+|\phi|^2)^2}|D_1\phi \pm \ri D_2\phi|^2\nn \\
&&+|D_1\psi \pm \ri D_2\psi|^2\pm\ri \left(\pa_1[\psi\overline{D_2 \psi}]-\pa_2[\psi\overline{D_1\psi}]\right)\nn \\
&&\pm (a+c\zeta)\hat{F}_{12}\pm (b+d\zeta)\tilde{F}_{12}\pm J_{12}(\phi).
\eea
which implies that the energy admits a lower bound
\be\lb{5.17}
E=\int_{\bfR^2} \mathcal{H}\, \dd x\ge\pm (a+c\zeta)\int_{\bfR^2} \hat{F}_{12}\, \dd x\pm(b+d\zeta)\int_{\bfR^2} \tilde{F}_{12}\, \dd x\pm \int_{\bfR^2}J_{12}(\phi)\,\dd x.
\ee
The energy lower bound is obtained if the quadratic terms in \eq{5.16} simultaneously vanish:
\bea
\hat{F}_{12}\pm \lm\left(a\frac{|\phi|^2-1}{|\phi|^2+1}+c(|\psi|^2-\zeta)\right)&=&0,\lb{5.18}\\
\tilde{F}_{12}\pm\frac{2G^2}{\ka^2}\left(b\frac{|\phi|^2-1}{|\phi|^2+1}+d(|\psi|^2-\zeta)\right)&=&0,\lb{5.19}\\
D_1\phi\pm \ri D_2 \phi &=& 0, \lb {5.20}\\
D_1\psi\pm \ri D_2 \psi &=& 0.\lb{5.21}
\eea
Rewriting \eq{5.18} and \eq{5.19}, we get
\bea
\hat{F}_{12}&=&\mp \lm\left(a\frac{|\phi|^2-1}{|\phi|^2+1}+c(|\psi|^2-\zeta)\right),\lb{5.22}\\
\tilde{F}_{12}&=&\mp \frac{2G^2}{\ka^2}\left(b\frac{|\phi|^2-1}{|\phi|^2+1}+d(|\psi|^2-\zeta)\right).\lb{5.23}
\eea

Let
\be\lb{5.24}
\phi: \left\{q'_1,\dots,q'_{M_1}\right\},\quad \left\{{p}'_1,\dots,{p}'_{N_1}\right\}; \quad \psi:\left\{{q}''_1,\dots,{q}''_{M_2}\right\}
\ee
be the prescribed zeros and poles of $\phi$ and the zeros of $\psi$, respectively.
Then, combining  \eq{3.23}, \eq{5.22}, \eq{5.23}, the system of equations \eq{5.18}-\eq{5.21} is reduced via $u=\ln|\phi|^2$ and $v=\ln|\psi|^2$
to
\bea
\Delta u&=&2a\lm\left(a\frac{\e^u-1}{\e^u+1}+c(\e^v-\zeta)\right)\nn \\
&&+\frac{4b}{\ka^2}\left(\frac{2b^2\e^u}{(1+\e^u)^2}+d^2\e^v\right)\left(b\frac{\e^u-1}{\e^u+1}+d(\e^v-\zeta)\right)
+4\pi\sum_{s=1}^{M_1} \delta_{q'_s}-4\pi\sum_{s=1}^{N_1} \delta_{p'_s},\nn \\
&&\lb{5.25}\\
\Delta v&=&2c\lm\left(a\frac{\e^u-1}{\e^u+1}+c(\e^v-\zeta)\right)\nn \\
&&+\frac{4d}{\ka^2}\left(\frac{2b^2\e^u}{(1+\e^u)^2}+d^2\e^v\right)\left(b\frac{\e^u-1}{\e^u+1}+d(\e^v-\zeta)\right)
+4\pi\sum_{s=1}^{M_2} \delta_{q''_s}.\lb{5.26}
\eea
With the vortex and antivortex points assigned in \eq{5.24}, we introduce the background function $v_0^1$ satisfying
\be\lb{5.27}
\Delta v_0^1=-\frac{4 \pi N_1}{|\Om|}+4 \pi \sum_{s=1}^{N_1} \delta_{p'_{s}}.
\ee
In view of \eq{3.26},  \eq{5.25}, \eq{5.26}, and the substitutions $u=u_0^1-v_0^1+w_1, v=u_0^2+w_2$, we obtain
\bea
\Delta w_1&=&2\lm a\left(a\frac{\e^{u_0^1-v_0^1+w_1}-1}{\e^{u_0^1-v_0^1+w_1}+1}+c(\e^{u_0^2+w_2}-\zeta)\right)\nn \\
&&+\frac{4b}{\ka^2}\left(\frac{2b^2\e^{u_0^1-v_0^1+w_1}}{(1+\e^{u_0^1-v_0^1+w_1})^2}+d^2\e^{u_0^2+w_2}\right)\left(b\frac{\e^{u_0^1-v_0^1+w_1}-1}{\e^{u_0^1-v_0^1+w_1}+1}
+d(\e^{u_0^2+w_2}-\zeta)\right)\nn \\
&&+\frac{4\pi(M_1-N_1)}{|\Om|},\lb{5.28}\\
\Delta w_2&=&2\lm c\left(a\frac{\e^{u_0^1-v_0^1+w_1}-1}{\e^{u_0^1-v_0^1+w_1}+1}+c(\e^{u_0^2+w_2}-\zeta)\right)\nn \\
&&+\frac{4d}{\ka^2}\left(\frac{2b^2\e^{u_0^1-v_0^1+w_1}}{(1+\e^{u_0^1-v_0^1+w_1})^2}+d^2\e^{u_0^2+w_2}\right)\left(b\frac{\e^{u_0^1-v_0^1+w_1}-1}{\e^{u_0^1-v_0^1+w_1}+1}
+d(\e^{u_0^2+w_2}-\zeta)\right)\nn \\
&&+\frac{4\pi M_2}{|\Om|},\lb{5.29}
\eea
which is now in a regular (singularity free) form.
Integrating \eq{5.28} and \eq{5.29}, we arrive at the formulas
\bea
a\lm\int f_3\,\dd x+\frac{2b}{\ka^2}\int G^2f_4\,\dd x&=&-2\pi(M_1-N_1),\lb{5.30}\\
c\lm\int f_3\,\dd x+\frac{2d}{\ka^2}\int G^2f_4\,\dd x&=&-2\pi M_2,\lb{5.31}
\eea
where $f_3$ and $f_4$ are defined as
\bea
{f}_3&=&f_3(u_0^1-v_0^1+w_1,u_0^2+w_2)=a\left(\frac{\e^{u_0^1-v_0^1+w_1}-1}{\e^{u_0^1-v_0^1+w_1}+1}\right)+c(\e^{u_0^2+w_2}-\zeta),\lb{5.32}\\
{f}_4&=&f_4(u_0^1-v_0^1+w_1,u_0^2+w_2)=b\left(\frac{\e^{u_0^1-v_0^1+w_1}-1}{\e^{u_0^1-v_0^1+w_1}+1}\right)+d(\e^{u_0^2+w_2}-\zeta).\lb{5.33}
\eea
In view of \eq{5.22}, \eq{5.23}, \eq{5.30} and \eq{5.31}, we derive the magnetic fluxes
\bea
\hat{\Phi}&=&\int \hat{F}_{12}\,\dd x=\mp\int \lm f_3\,\dd x=\pm\frac{2\pi(d (M_1-N_1)-b M_2)}{ad-bc},\lb{5.34}\\
\tilde{\Phi}&=&\int\tilde{F}_{12}\,\dd x=\mp\int\frac{2}{\ka^2}G^2 f_4\,\dd x=\pm\frac{2\pi(a M_2-c (M_1-N_1))}{ad-bc}.\lb{5.35}
\eea
Besides, from \cite{SSY,XY}, we have the Thom class following the expression:
\be\lb{5.36}
\frac{1}{2\pi}\int J_{12}(\phi)\,\dd x=\pm 2 N_1.
\ee
Thus, inserting \eq{5.34}-\eq{5.36} into \eq{5.17}, we arrive at
\be\lb{5.37}
E=2\pi (M_1+N_1+\zeta M_2),
\ee
which is different from energy of the pure Chern-Simons model \eq{2.14}, where the latter depends only on the vortex numbers $M_i$.
With \eq{5.6}, we obtain the electric charge of the system \eq{5.1}
\be\lb{5.38}
Q=\int\ka\tilde{F}_{12}\, \dd x=\pm\frac{2\ka\pi(a M_2-c (M_1-N_1))}{ad-bc}.
\ee
In contrast to \eq{3.36}, we see from \eq{5.38} that, for the vortex-antivortex system, the total charge, not only depends on the vortex number $M_i$, but also on the antivortex number $N_1$.

\section{Bogomol'nyi equations for Chern-Simons vortices coupled with Born--Infeld vortices and antivortices}\lb{sec6}
In Section \ref{sec4}, we obtained the Bogomol'nyi equations for Chern-Simons and Born-Infeld vortices. In order to extend our investigation,  here we construct a model that accommodates both Chern-Simons vortices and Born-Infeld vortices-antivortices, whose static Hamiltonian energy density reads
\bea\lb{6.7}
\mathcal{H}&=&b_1^2\left(\sqrt{1+\frac{1}{b_1^2}\hat{F_{12}^2}}-1\right)+\left(\frac{2b^2|\phi|^2}{(1+|\phi|^2)^2}+d^2|\psi|^2\right)\tilde{A}_0^2+\frac{2}{(1+|\phi|^2)^2}|D_i\phi|^2\nn \\
&&+|D_i\psi|^2+V.
\eea
Comparing \eq{5.7} with \eq{6.7}, we observe that the Maxwell action in \eq{5.7} is replaced by the Born-Infeld form in \eq{6.7}.
The variational principle yields the Euler-Lagrange equations
\bea
D_i\left(\frac{D_i\phi}{(1+|\phi|^2)^2}\right)&=&-\frac{2|D_i\phi|^2\phi}{(1+|\phi|^2)^3}+\frac12\frac{\pa V}{\pa{\overline{\phi}}}+\frac{2b^2\tilde{A}_0^2|\phi|^2\phi}{(1+|\phi|^2)^3}-\frac{b^2\tilde{A}_0^2\phi}{(1+|\phi|^2)^2},\lb{6.8}  \\
D^2_i\psi&=&\frac{\pa V(|\phi|^2,|\psi|^2)}{\pa{\overline{\psi}}}-d^2\tilde{A}_0^2\psi, \lb{6.9} \\
\pa_j \left(\frac{\hat{F}_{ij}}{\sqrt{1+\frac{1}{b_1^2}\hat{F}_{12}^2}}\right)&=&\frac{2a\ri}{(1+|\phi|^2)^2}[\phi\overline{D_i\phi}-\overline{\phi}D_i\phi])+ \ri c[\psi\overline{D_i\psi}-\overline{\psi}D_i\psi],\lb{6.10}\\
\ka\vep_{ij}\pa_j\tilde{A}_0&=&\frac{2\ri b}{(1+|\phi|^2)^2}[\phi\overline{D_i\phi}-\overline{\phi}D_i\phi])+ \ri d[\psi\overline{D_i\psi}-\overline{\psi}D_i\psi].\lb{6.11}
\eea
Furthermore, the Gauss Law \eq{5.6} also holds in this model.

Following the same steps as in Section 5, together with \eq{3.13} and \eq{5.14a}, we regroup the Hamiltonian density \eq{6.7} as
\bea\lb{6.13}
\cal{H}&=&\frac{1}{2}\left(\hat{F}_{12}\pm F\left(a\frac{|\phi|^2-1}{|\phi|^2+1}+c(|\psi|^2-\zeta)\right)\right)^2F^{-1}+\frac{b_1^2}{2}(FU-1)^2F^{-1}\nn \\
&&-b_1^2+b_1^2U\mp\hat{F}_{12}\left(a\frac{|\phi|^2-1}{|\phi|^2+1}+c(|\psi|^2-\zeta)\right) \nn \\
&&+\left(\frac{\ka\tilde{F}_{12}}{2G}\pm\frac{G}{\ka}\left(b\frac{|\phi|^2-1}{|\phi|^2+1}+d(|\psi|^2-\zeta)\right)\right)^2\mp \tilde{F}_{12}\left(b\frac{|\phi|^2-1}{|\phi|^2+1}+d(|\psi|^2-\zeta)\right)\nn \\
&&-\frac{G^2}{\ka^2}\left(b\frac{|\phi|^2-1}{|\phi|^2+1}+d(|\psi|^2-\zeta)\right)^2\nn \\
&&+\frac{2}{(1+|\phi|^2)^2}\left(|D_1\phi \pm \ri D_2\phi|^2\pm \ri (D_1\phi \overline{D_2\phi}-\overline{D_1\phi}D_{2}\phi)\right)\nn \\
&&+|D_1\psi \pm \ri D_2\psi|^2 \pm \ri (\pa_1[\psi \overline{D_2\psi}]-\pa_2[\psi \overline{D_1 \psi}])\pm (c\hat{F}_{12}+d\tilde{F}_{12})|\psi|^2+V,
\eea
where $G$ is defined as \eq{5.8} and $F$ and $U$ are given  by
\be\lb{6.14}
F={\sqrt{1+\frac1{b_1^2}\hat{F}_{12}^2}},\quad U=\sqrt{1-\frac{1}{b_1^2}\left(a\frac{|\phi|^2-1}{|\phi|^2+1}+c(|\psi|^2-\zeta)\right)},
\ee
respectively. The desired Bogomol'nyi structure motivates us to choose the potential density as
\bea\lb{6.15}
V&=&b_1^2\left(1-\sqrt{1-\frac{1}{b_1^2}\left(a\frac{|\phi|^2-1}{|\phi|^2+1}+c(|\psi|^2-\zeta)\right)^2}\right)\nn \\
&&+\frac{1}{\ka^2}\left(\frac{2b^2|\phi|^2}{(1+|\phi|^2)^2}+d^2|\psi|^2\right)\left(b\frac{|\phi|^2-1}{|\phi|^2+1}+d(|\psi|^2-\zeta)\right).
\eea
Inserting \eq{5.14} and \eq{6.15} into \eq{6.13}, we have
\bea\lb{6.16}
\cal{H}&=&\frac{1}{2}\left(\hat{F}_{12}\pm F\left(a\frac{|\phi|^2-1}{|\phi|^2+1}+c(|\psi|^2-\zeta)\right)\right)^2F^{-1}+\frac{b_1^2}{2}(FU-1)^2F^{-1}\nn \\
&&+\left(\frac{\ka\tilde{F}_{12}}{2G}\pm\frac{G}{\ka}\left(b\frac{|\phi|^2-1}{|\phi|^2+1}+d(|\psi|^2-\zeta)\right)\right)^2 \nn \\
&&+\frac{2}{(1+|\phi|^2)^2}|D_1\phi \pm \ri D_2\phi|^2+|D_1\psi \pm \ri D_2\psi|^2 \pm \ri (\pa_1[\psi \overline{D_2\psi}]-\pa_2[\psi \overline{D_1 \psi}])\nn \\
&&\pm (a+c\zeta)\hat{F}_{12}\pm (b+d\zeta)\tilde{F}_{12}\pm J_{12}(\phi),
\eea
Thus we obtain the same energy expression as \eq{5.17}, which is saturated when the self-dual or anti-self-dual equations hold:
\bea
\hat{F}_{12}\pm F\left(a\frac{|\phi|^2-1}{|\phi|^2+1}+c(|\psi|^2-\zeta)\right)&=&0,\lb{6.18}\\
FU-1&=&0, \lb{6.19}\\
\frac{\ka\tilde{F}_{12}}{2G}\pm\frac{G}{\ka}\left(b\frac{|\phi|^2-1}{|\phi|^2+1}+d(|\psi|^2-\zeta)\right)&=& 0, \lb{6.20} \\
D_1\phi\pm \ri D_2 \phi &=& 0, \lb {6.21}\\
D_1\psi\pm \ri D_2 \psi &=& 0. \lb{6.22}
\eea
These are the Bogomol'nyi equations of the model \eq{6.7}. Since \eq{6.18} and \eq{6.19} are not independent, they may be reduced to a single equation
\be\lb{6.23}
\hat{F}_{12}=\mp \frac{a\frac{|\phi|^2-1}{|\phi|^2+1}+c(|\psi|^2-\zeta)}{\sqrt{1-\frac{1}{b_1^2}\left(a\frac{|\phi|^2-1}{|\phi|^2+1}+c(|\psi|^2-\zeta)\right)^2}},\\
\ee
For the prescribed points \eq{5.24}, we combine \eq{3.23}, \eq{6.20} and \eq{6.23} to derive the governing equations
\bea
\Delta u&=&\frac{2a\left(a\frac{\e^u-1}{\e^u+1}+c(\e^v-\zeta)\right)}{\sqrt{1-\frac{1}{b_1^2}\left(a\frac{\e^u-1}{\e^u+1}+c(\e^v-\zeta)\right)^2}}\nn \\
&&+\frac{4b}{\ka^2}\left(\frac{2b^2\e^u}{(1+\e^u)^2}+d^2\e^v\right)\left(b\frac{\e^u-1}{\e^u+1}+d(\e^v-\zeta)\right)
+4\pi\sum_{s=1}^{M_1} \delta_{q'_s}-4\pi\sum_{s=1}^{N_1} \delta_{p'_s},\nn \\
&&\lb{6.24}\\
\Delta v&=&\frac{2c\left(a\frac{\e^u-1}{\e^u+1}+c(\e^v-\zeta)\right)}{\sqrt{1-\frac{1}{b_1^2}\left(a\frac{\e^u-1}{\e^u+1}+c(\e^v-\zeta)\right)^2}}\nn \\
&&+\frac{4d}{\ka^2}\left(\frac{2b^2\e^u}{(1+\e^u)^2}+d^2\e^v\right)\left(b\frac{\e^u-1}{\e^u+1}+d(\e^v-\zeta)\right)
+4\pi\sum_{s=1}^{M_2} \delta_{q''_s},\lb{6.25}
\eea
where $u=\ln|\phi|^2$ and $v=\ln|\psi|^2$.
Integrating \eq{6.24} and \eq{6.25} yields the equations
\bea
a\int\frac{f_3}{\sqrt{1-\frac{1}{b_1^2}f_3^2}}\,\dd x+\frac{2b}{\ka^2}\int G^2 f_4\,\dd x&=& -2\pi (M_1-N_1),\lb{6.26}
\\
c\int\frac{f_3}{\sqrt{1-\frac{1}{b_1^2}f_3^2}}\,\dd x+\frac{2d}{\ka^2}\int G^2 f_4\,\dd x&=& -2\pi M_2,\lb{6.27}
\eea
where $f_3=f_3(u,v)$ and $f_4=f_4(u,v)$ are defined by \eq{5.32}-\eq{5.33} with $u=u_0^1-v_0^1+w_1$ and $v=u_0^2+w_2$.

By simple calculation on \eq{6.26} and \eq{6.27}, we derived from \eq{6.20} and \eq{6.23} the magnetic fluxes
\bea
\hat{\Phi}&=&\int \hat{F}_{12}\,\dd x=\mp\int \frac{f_3}{\sqrt{1-\frac{1}{b_1^2}f_3^2}}\,\dd x=\pm\frac{2\pi(d (M_1-N_1)-b M_2)}{ad-bc},\lb{6.28}\\
\tilde{\Phi}&=&\int\tilde{F}_{12}\,\dd x=\mp\int\frac{2}{\ka^2}G^2 f_4\,\dd x=\pm\frac{2\pi(a M_2-c (M_1-N_1))}{ad-bc},\lb{6.29}
\eea
which share the same expressions as \eq{5.34} and \eq{5.35}. As a consequence, in view of \eq{5.6} and \eq{5.17}, we obtain \eq{5.37} and \eq{5.38} again, respectively.

In contrast to the pure Chern-Simons theory \eq{2.9}, vortices and antivortices can coexist in our mixed models \eq{5.7} and \eq{6.7}, a property that is difficult to achieve in conventional Chern-Simons theories, even in the simplest single-species case \eq{2.1}. Meanwhile, these new models inherit both the dyonic property of the Chern-Simons system and the vortex-antivortex coexistence of the harmonic map model, owing to a strong interplay between the two sectors, as can be seen from the potential density \eq{5.16} and \eq{6.15}. These mixed models provide a platform for studying a rich spectrum of soliton solutions.
\section{Bradlow bounds}\lb{sec7}
In this section, we establish some necessary conditions for existence of solutions for the systems \eq{3.7}, \eq{4.7}, \eq{5.7}, and \eq{6.7} in the doubly-periodic domain: for the vortex-only systems \eq{3.7} and \eq{4.7}, the vortex numbers $M_i$ are constrained by a bound, called Bradlow bounds \cite{NM,SBG}; for the vortex-antivortex system \eq{5.7} and \eq{6.7}, this bound is imposed on both the difference  $|M_1-N_1|$ and $M_2$.
In this discussion, the value of the charge parameters $a,b,c,d$ will naturally and technically play subtle roles. For the sake of simplicity, we shall assume the positive condition
\be\lb{7.1}
a,b,c,d>0.
\ee
From \eq{3.0} and \eq{3.0a}, we know that by flipping the sign of either $\hat{A}_i \mapsto -\hat{A}_i$ or $\tilde{A}_i \mapsto -\tilde{A}_i$, or both, the condition \eq{7.1} can cover all possible sign configurations of the charge parameters:
\be
(a,c,b,d)\sim (--++),(++--),(----).
\ee
\subsection{The vortex model \eq{3.7}}
This model accommodates prescribed vortices that are realized by the zeros of the Higgs fields $\phi$ and $\psi$ as described by \eq{3.22} and governed by the equations \eq{3.24} and \eq{3.25}. In view of \eq{3.29}, we obtain the necessary condition
\be\lb{7.2}
\frac{2\pi(b M_2-d M_1)}{\lm (ad-bc)}+(a \xi +c\zeta)|\Om|=\int_{\Om}(a \e^{u_0^1+w_1}+c\e^{u_0^2+w_2})\,\dd x>0,
\ee
through the positive condition \eq{7.1}.
On the other hand,  the integrand on the left-hand side of \eq{3.30} can be rewritten as
\bea\lb{7.3}
\frac{2\pi (c M_1-a M_2)\ka^2}{ad-bc}&=&\int_{\Om} \left\{2b^2\e^u[b(\e^u-\xi)-d\zeta]+2d^2\e^{v}[d(\e^v-\zeta)-b\xi]\right\}\,\dd x\nn \\
&&+\int_{\Om} 2bd(b+d)\e^{u+v}\,\dd x \nn \\
&=&\int_{\Om} \left\{2b^3\left(\e^u-\frac{\xi+\frac{d}{b}\zeta}{2}\right)^2+2d^3\left(\e^v-\frac{\zeta+\frac{b}{d}\xi}{2}\right)^2\right\}\,\dd x \nn \\
&&+\int_{\Om} \left\{2bd(b+d)\e^{u+v}- \frac{b^3(\xi+\frac{d}{b}\zeta)^2}{2}-\frac{d^3(\zeta+\frac{b}{d}\xi)^2}{2}\right\}\,\dd x,\nn \\
\eea
where we set $u=u_0^1+w_1$  and $v=u_0^2+w_2$ to save space. It is obvious that, under the condition \eq{7.1}, the identity \eq{7.3} gives rise to the following inequality:
\be\lb{7.4}
\frac{2\pi (c M_1-a M_2)\ka^2}{ad-bc}+ \left(\frac{b^3(\xi+\frac{d}{b}\zeta)^2}{2}+\frac{d^3(\zeta+\frac{b}{d}\xi)^2}{2}\right)|\Om|>0.
\ee
As a result, the inequality constraints \eq{7.2} and \eq{7.4} yield the Bradlow bounds
\bea
M_1&<&\frac{|\Om|}{2\pi}\bigg[\frac{b}{\ka^2}\left(\frac{b^3}{2}(\xi+\frac{d}{b}\zeta)^2+\frac{d^3}{2}(\zeta+\frac{b}{d}\xi)^2\right)+\lm a(a\xi+c\zeta)\bigg],\lb{7.5} \\
M_2&<&\frac{|\Om|}{2\pi}\bigg[\frac{d}{\ka^2}\left(\frac{b^3}{2}(\xi+\frac{d}{b}\zeta)^2+\frac{d^3}{2}(\zeta+\frac{b}{d}\xi)^2\right)+\lm c(a\xi+c\zeta)\bigg].\lb{7.6a}
\eea
The topological-geometric constraints exert the upper bounds on the vortex number $M_1$ and $M_2$, showing that they cannot be arbitrarily large in the compact domain.
\subsection{The vortex model \eq{4.7}}
In this subsection, we calculate the Bradlow bounds for a system of Born--Infeld and Chern--Simons vortices given by the model \eq{4.7}, which are represented by the zeros of the Higgs fields $\phi$ and $\psi$, respectively, as stated in \eq{3.22}. Since the expression \eq{7.3} remains valid in this case, here we omit the redundant steps and directly use \eq{7.4}. With the background functions $u_0^1,u_0^2$ defined in \eq{3.26}, together with the substitutions $u=u_0^1+w_1$ and $v=u_0^2+w_2$, we rewrite the integrand on the left-hand side of \eq{4.25} as
\be\lb{7.6aa}
h(t_1,t_2)=\frac{a(t_1-\xi)+c(t_2-\zeta)}{\sqrt{1-\frac{1}{b_1^2}[a(t_1-\xi)+c(t_2-\zeta))]^2}},
\ee
where $t_1=\e^{u_0^1+w_1}$ and $t_2=\e^{u_0^2+w_2}$.
On the other hand, a simple calculation shows that the partial derivatives
\be\lb{7.7}
\left(\frac{\pa}{\pa t_1},\frac{\pa}{\pa t_2}\right)h=\frac{1}{(1-\frac{1}{b_1^2}\left(a(t_1-\xi)+c(t_2-\zeta)\right)^2)^{\frac32}}(a,c),
\ee
are all positive. Thus, according to \eq{4.25} and \eq{7.7}, we have
\be\lb{7.7a}
\frac{2\pi(b M_2-d M_1)}{ad-bc}=\int_{\Om} h(t_1,t_2)\, \dd x>\int_{\Om} h(0,0)\,\dd x=\frac{-(a\xi+c\zeta)|\Om|}{\sqrt{1-\frac{1}{b_1^2}(a\xi+c\zeta)^2}}.
\ee
Combining \eq{7.4} and \eq{7.7a}, we found that $M_1$ and $M_2$ enjoy the Bradlow bounds
\bea
M_1&<&\frac{|\Om|}{2\pi}\bigg[\frac{b}{\ka^2}\left(\frac{b^3}{2}(\xi+\frac{d}{b}\zeta)^2+\frac{d^3}{2}(\zeta+\frac{b}{d}\xi)^2\right)
+\frac{a(a\xi+c\zeta)}{\sqrt{1-\frac{1}{b_1^2}(a\xi+c\zeta)^2}}\bigg],\\
M_2&<&\frac{|\Om|}{2\pi}\bigg[\frac{d}{\ka^2}\left(\frac{b^3}{2}(\xi+\frac{d}{b}\zeta)^2+\frac{d^3}{2}(\zeta+\frac{b}{d}\xi)^2\right)
+\frac{c(a\xi+c\zeta)}{\sqrt{1-\frac{1}{b_1^2}(a\xi+c\zeta)^2}}\bigg],
\eea
which are Bradlow-type bounds and similar to \eq{7.5} and \eq{7.6a}.
\subsection{The vortex and antivortex model \eq{5.7}}
We now establish the Bradlow bounds for the mixed vortex-antivortex model \eq{5.7} which accommodates $M_1$ vortices and $N_1$ antivortice (represented by the zeros and poles of $\phi$) and $M_2$ vortices (represented by the zeros of $\psi$), as indicated in \eq{5.24}.  Solving \eq{5.30} and \eq{5.31}, we easily obtain
\be\lb{7.8}
\frac{2\pi(b M_2-d(M_1-N_1))}{\lm (ad-bc)}=\int_{\Om} \left(a\frac{\e^{u_0^1-v_0^1+w_1}-1}{\e^{u_0^1-v_0^1+w_1}+1}+c(\e^{u_0^2+w_2}-\zeta)\right)\,\dd x.
\ee
With $t_1=\e^{u_0^1-v_0^1+w_1}$ and $t_2=\e^{u_0^2+w_2}$, we rewrite the integrand on the right hand side of \eq{7.8} as
\be
g(t_1,t_2)=a\frac{t_1-1}{t_1+1}+c(t_2-\zeta),
\ee
whose partial derivative are all positive.
Substituting this into \eq{7.8} yields
\be\lb{7.10}
\frac{2\pi(b M_2-d(M_1-N_1))}{\lm (ad-bc)}>-(a+c\zeta)|\Om|.
\ee
On the other hand, from \eq{5.30} and \eq{5.31}, we also have, with $u=u_0^1-v_0^1+w_1$ and $v=u_0^2+w_2$,
\bea
\frac{\pi(c(M_1-N_1)-a M_2)\ka^2}{ad-bc}&=&\int_{\Om}\left(\frac{2b^2\e^u}{(1+\e^u)^2}+d^2\e^v\right)\left(b\frac{\e^u-1}{\e^u+1}+d(\e^v-\zeta)\right)\,\dd x \nn \\
&=&\int_{\Om}\left\{\frac{2b^2(b-d\zeta)}{(1+\e^u)^3}\left(\e^u-\frac{b+d\zeta}{2(b-d\zeta)}\right)^2+\frac{2b^2d\e^{u+v}}{(1+\e^u)^2}\right\}\, \dd x \nn \\
&&-\int_{\Om}\left\{\frac{b^2(b+d\zeta)^2}{2(1+\e^u)^3(b-d\zeta)}-\frac{2bd^2\e^{u+v}}{1+\e^u}\right\}\,\dd x\nn \\
&&+\int_{\Om} \left\{ d^3\left(\e^v-\frac{b+d\zeta}{2d}\right)^2-\frac{d(b+d\zeta)^2}{4}\right\}\,\dd x, \nn
\eea
which implies that, if $b-d\zeta>0$, the following inequality constraint holds:
\be\lb{7.11}
\frac{\pi(c(M_1-N_1)-a M_2)\ka^2}{ad-bc}+\int_{\Om}\frac{b^2(b+d\zeta)^2}{2(1+\e^u)^3(b-d\zeta)}\,\dd x+\int_{\Om}\frac{d(b+d\zeta)^2}{4}\,\dd x>0.
\ee
Thus \eq{7.11} leads to the necessary condition
\be\lb{7.12}
\frac{\pi(c(M_1-N_1)-a M_2)\ka^2}{ad-bc}+\frac{d(b+d\zeta)^2}{4}|\Om|>-\frac{b^2(b+d\zeta)^2|\Om|}{2(b-d\zeta)}.
\ee
In view of \eq{7.10} and \eq{7.12}, we arrive at the Bradlow bounds
\bea
M_1-N_1&<&\frac{|\Om|}{2\pi}\left(\frac{b}{\ka^2}\left(\frac{d(b+d\zeta)^2}{2}+\frac{b^2(b+d\zeta)^2}{(b-d\zeta)}\right)+\lm a(a+c\zeta)\right)\equiv K,\lb{7.13} \\
M_2&<&\frac{|\Om|}{2\pi}\left(\frac{d}{\ka^2}\left(\frac{d(b+d\zeta)^2}{2}+\frac{b^2(b+d\zeta)^2}{(b-d\zeta)}\right)+\lm c(a+c\zeta)\right).\lb{7.14}
\eea
The inequality \eq{7.13} provides an upper bound on the difference of the numbers of vortices $M_1$ and antivortices $N_1$, rather than on $M_1$ or $N_1$ individually. In other words, $M_1$  and $N_1$ may be arbitrarily large as long as their difference remains bounded. This contrasts with the vortex-only models \eq{3.7} and \eq{4.7}, where the vortex number itself is bounded from above.

In order to derive a lower bound of $M_1-N_1$, we consider the transformation
\be\lb{7.12a}
\phi\mapsto \frac1{\phi},\quad \psi\mapsto \psi; \quad \hat{A}_i\mapsto -\hat{A}_i,\quad \tilde{A}_i\mapsto -\tilde{A}_i; \quad (a,b)\mapsto (a,b),\quad (c,d)\mapsto -(c,d).
\ee
which renders both the model \eq{5.7} with the potential density function \eq{a5.16} and its governing equations invariant.
Exchanging the roles of the zeros and poles of $\phi$ yields a solution where the field $\phi$ has $N_1$ zeros ${p}'_1,\dots,{p}'_{N_1}$ and $M_1$ poles
${q}'_1,\dots,{q}'_{M_1}$, with multiplicities counted as in \eq{5.24}. Thus \eq{7.13} results in $N_1-M_1<K$.
In view of this inequality constraint and \eq{7.13}, we obtain $|M_1-N_1|<K$ or
\be\lb{7.15}
|M_1-N_1|<\frac{|\Om|}{2\pi}\left(\frac{b}{\ka^2}\left(\frac{d(b+d\zeta)^2}{2}+\frac{b^2(b+d\zeta)^2}{(b-d\zeta)}\right)+\lm a(a+c\zeta)\right).
\ee
The bounds \eq{7.14} and \eq{7.15} are the Bradlow bounds for the model \eq{5.1} which is a system of $M_1$ vortices and $N_1$ antivortices of the Maxwell type and $M_2$ vortices of the Chern-Simons type over a lattice cell domain $\Om$.
\subsection{The vortex and antivortex model \eq{6.7}}
By simple algebraic manipulation on \eq{6.26} and \eq{6.27}, we obtain the same result as \eq{7.12}. In other words, here we only need to deal with the term associated with Born-Infeld, that is
\be\lb{7.16}
\frac{2\pi(b M_2-d(M_1-N_1))}{ad-bc}=\int_{\Om}\frac{f_3}{\sqrt{1-\frac{1}{b_1^2}f_3^2}}\,\dd x.
\ee
Using the background function $u_0^1,u_0^2$ and $v_0^1$ defined in \eq{3.26} and \eq{5.27}, respectively, and the substitutions $u=u_0^1-v_0^1+w_1$ and $v=u_0^2+w_2$, we rewrite the integrand of the right hand side of \eq{7.16} as
\be\lb{7.16a}
h(t_1,t_2)=\frac{a\frac{t_1-1}{t_1+1}+c(t_2-\zeta)}{\sqrt{1-\frac{1}{b_1^2}\left(a\frac{t_1-1}{t_1+1}+c(t_2-\zeta)\right)^2}}>h(0,0),
\ee
where $t_1=\e^{u_0^1-v_0^1+w_1}$ and $t_2=\e^{u_0^2+w_2}$.
Substituting \eq{7.16a} into \eq{7.16} leads to
\be\lb{7.17}
\frac{2\pi(b M_2-d(M_1-N_1))}{ad-bc}=\int_{\Om} h(t_1,t_2)\,\dd x>-\frac{(a+c\zeta)|\Om|}{\sqrt{1-\frac{1}{b_1^2}(a+c\zeta)^2}}.
\ee
From this inequality together with \eq{7.12} and \eq{7.12a}, we again obtain the Bradlow bounds
\bea
|M_1-N_1|&<&\frac{|\Om|}{2\pi}\left(\frac{b}{\ka^2}\left(\frac{d(b+d\zeta)^2}{2}+\frac{b^2(b+d\zeta)^2}{(b-d\zeta)}\right)+\frac{a(a+c\zeta)}{\sqrt{1-\frac{1}{b_1^2}(a+c\zeta)^2}}\right), \\
M_2&<&\frac{|\Om|}{2\pi}\left(\frac{d}{\ka^2}\left(\frac{d(b+d\zeta)^2}{2}+\frac{b^2(b+d\zeta)^2}{(b-d\zeta)}\right)+\frac{c(a+c\zeta)}{\sqrt{1-\frac{1}{b_1^2}(a+c\zeta)^2}}\right),
\eea
which are similar to \eq{7.14} and \eq{7.15}.
\section{Conclusions}\lb{sec8}
In this paper we have obtained several new Bogomol'nyi equations from a broad class of Abelian product gauge field theories in which one of the two $U(1)$ sector is governed by a Chern-Simons-Higgs Lagrangian.
\begin{itemize}
    \item We demonstrated that the product Abelian Chern-Simons-Higgs theory \eq{2.9} of Han and Yang \cite{HY0}, considered a general model in which two Higgs scalar fields $\phi$ and $\psi$ carry charges $(a,b)$ and $(c,d)$ for arbitrary real parameters $a,b,c,d$, respectively, can be extended into two hybrid models in which one of the gauge fields is replaced by Maxwell term (or Born-Infeld nonlinear electrodynamics), and that such new models also enjoy BPS structure as an important reduction of second-order equations of motion. (Sections \ref{sec3} and \ref{sec4}).
    \item We demonstrated that the mixed Chern-Simons systems allow the vortex and anivortex to coexist by modifying one of the Chern-Simons potential into either harmonic map potential or radical-root nonlinear potential, which are governed by Maxwell theory and Born-Infeld nonlinear electrodynamics, respectively. (Sections \ref{sec5} and \ref{sec6})
    \item For each model, we obtained corresponding Bogomol'nyi equations whose solutions allow the energy functional to attain a topological lower bound. We also calculated the composite energy, magnetic flux and total charge. These physical quantities are quantized: the energy depends only on the vortex and antivortex numbers, while the magnetic flux and total charge on both these numbers and the charges $a,b,c,d$. Moreover, our hybrid vortex-only systems \eq{3.7} and \eq{4.7} take the same values for these physical quantities as the pure Chern-Simons system \eq{2.9}. However, in our mixed vortex-antivortex systems \eq{5.7} and \eq{6.7}, these quantities are different from those in the system \eq{2.9} and depend on both the vortex numbers $M_i$ and the antivortex numbers $N_1$.
    \item A key finding of this paper is that the results are different between the full-plane and a doubly-periodic domain. Due to topological-geometric constraints on a compact domain, the models \eq{3.1}, \eq{4.7}, \eq{5.7} and \eq{6.7} are subject to Bradlow-type bounds. For the first two models (vortex-only systems), the Bradlow bounds are imposed on the vortex numbers $M_i$. For the last two models (vortex-antivortex systems), the bounds are imposed on both the vortex number $M_2$ and the difference between vortex number $M_1$ and antivortex number $N_1$. These bounds have a direct influence on the energy: for vortex-vortex systems, the Bradlow bounds lead to a finite energy spectrum. In contrast, for vortex-antivortex systems, the energy is unbounded, since the numbers $M_1$ and $N_1$ may increase without bound while $|M_1-N_1|$ remains bounded.
\end{itemize}

In summary, the mixed $U(1)\times U(1)$ Chern-Simons models developed in this work provide a unified framework that bridges two different extensions. When our mixed models are regarded as a generalization of product Abelian Higgs, Born-Infeld or harmonic map models in which only magnetic charge is carried, the appearance of a Chern-Simons term brings dyonic degrees of freedom. On the other hand, when these models are viewed as an extension of product Chern-Simons models, the replacement of a Chern-Simons term by a Maxwell or Born-Infeld term leads to abundant BPS structures and naturally accommodates vortex-antivortex pairs by choosing suitable potentials. These two complementary perspectives exhibit the versatility of our mixed models. This work significantly enriches the zoo of vortex equations \cite{NM,SBG}.

\medskip

{\bf Declaration of interests.} I declare I have no competing interests.

{\bf Data access.} This article has no additional data.


\begin{thebibliography}{99}
\bibitem{AA}
A.A. Abrikosov, On the magnetic properties of superconductors of the second group,
{\em Sov. Phys} {\em JETP} {\bf 5} (1957) 1174-1182.

\bibitem{HP}
H.B. Nielsen, P. Olesen, Vortex line models for dual strings, {\em Nucl. Phys.} B {\bf 61} (1973) 45-61.

\bibitem{Y2}
Y. Yang, Coexistence of vortices and anti-vortices in an Abelian gauge theory, {\em Phys. Rev. Lett.} {\bf80} (1998) 26--29.

\bibitem{XY2}
A. Xu, Y. Yang, Bogomol'nyi Equations in Two-Species Born--Infeld Theories Governing Vortices and Antivortices, (2026) arXiv:2601.09091




\bibitem{Bo}
E.B. Bogomol'nyi, The stability of classical solutions, {\em Sov. J. Nucl. Phys.} {\bf 24} (1976) 449--454.


\bibitem{PS}
M.K. Prasad and C. M. Sommerfield, Exact classical solutions for the 't Hooft monopole and the Julia--Zee dyon, {\em Phys. Rev. Lett.} {\bf35} (1975) 760--762.


\bibitem{AO1}
 J. Ambjorn and P. Olesen, Anti-screening of large magnetic fields
by vector bosons, {\em Phys. Lett.} B {\bf214} (1988) 565--569.

\bibitem{AO2}
J. Ambjorn and P. Olesen, On electroweak magnetism, {\em Nucl. Phys.}
B {\bf315} (1989) 606--614.

\bibitem{AO3}
 J. Ambjorn and P. Olesen, A magnetic condensate solution of the
classical electroweak theory, {\em Phys. Lett.} B {\bf218} (1989)  67--71.

\bibitem{AO4}
 J. Ambjorn and P. Olesen, A condensate solution of the classical
electroweak theory which interpolates between the broken and the
symmetric phase, {\em Nucl. Phys.} B {\bf330} (1990) 193--204.

\bibitem{SYew1}
 J. Spruck and Y. Yang, On multivortices in the electroweak theory
I: Existence of periodic solutions, {\em Commun. Math. Phys.} {\bf 144} (1992) 1--16.

\bibitem{SYew2}
 J. Spruck and Y. Yang, On multivortices in the electroweak theory
II: Existence of Bogomol'nyi solutions in $\bfR^2$, {\em Commun. Math. Phys.}
{\bf 144} (1992) 215--234.

\bibitem{BL1}
G. Bimonte and G. Lozano, $Z$ flux-line lattices and self-dual equations in the standard model, {\em Phys. Rev.} D {\bf50} (1994) 6046--6050.

\bibitem{BL2}
G. Bimonte and G. Lozano, Vortex solutions in two-Higgs-doublet systems, {\em Phys. Lett.} B {\bf326} (1994) 270--275.

\bibitem{Yew}
Y. Yang, Topological solitons in the Weinberg--Salam theory, {\em Physica} D {\bf101} (1997) 55--94.

\bibitem{HKP}
J. Hong, Y. Kim and P.-Y. Pac,  Multivortex solutions of the Abelian Chern--Simons--Higgs
theory, {\em Phys. Rev. Lett.} \textbf{64} (1990) 2330--2333.

\bibitem{JW}
R. Jackiw and E. J. Weinberg,  Self-dual Chern--Simons vortices,  {\em Phys. Rev. Lett.}
\textbf{64} (1990) 2334--2337.

\bibitem{Caff-Y}
L.A. Caffarelli and Y. Yang, Vortex condensation in the Chern--Simons Higgs model: An
existence theorem, {\em Commun. Math. Phys.} {\bf168} (1995) 321--336.

\bibitem{Han-L-Y}
X. Han, C. S. Lin, and Y. Yang,
 Resolution of Chern--Simons--Higgs vortex equations, {\em Commun.  Math. Phys.} {\bf343} (2016) 701--724.

\bibitem{Han-Y2}
X. Han and Y. Yang, Doubly periodic solutions of
relativistic Chern--Simons--Higgs vortex equations, {\em Trans. Amer. Math. Soc.} {\bf368} (2016) 3565--3590.


\bibitem{V}
A. Vilenkin, Cosmic strings and domain walls, {\em Phys. Rep.} {\bf121} (1985) 263--315.

\bibitem{CG}
A. Comtet and G. W. Gibbons, Bogomol'nyi bounds for cosmic strings, {\em Nucl. Phys.} B {\bf299} (1988) 719--733.

\bibitem{Ycosmic1}
Y. Yang, Obstructions to the existence of static cosmic strings in an Abelian Higgs model, {\em Phys. Rev. Lett.} {\bf73} (1994) 10-13.

\bibitem{Ycosmic2}
Y. Yang, Prescribing topological defects for the coupled Einstein and Abelian Higgs equations, {\em Commun. Math. Phys.} {\bf170} (1995) 541--582.

\bibitem{Ybook}
Y. Yang, {\em Solitons in Field Theory and Nonlinear Analysis}, Springer Verlag, Berlin and New York, 2001.


\bibitem{BI1}
M. Born and L. Infeld, Foundation of the new field theory, {\em Nature} {\bf132} (1933) 1004.

\bibitem{BI2}
M. Born and L. Infeld, Foundation of the new field theory, {\em Proc. Roy. Soc.}
A {\bf 144} (1934) 425--451.

\bibitem{SH}
M. Shiraishi and S. Hirenzaki, Bogomol'nyi equations for vortices in Born--Infeld Higgs systems, {\em Inter. J. Mod. Phys.} A {\bf6} (1991) 2635--2647.

\bibitem{Ybi}
Y. Yang, Classical solutions in the Born--Infeld theory, {\em Proc.
Roy. Soc.}  A {\bf456} (2000) 615--640.

\bibitem{Han}
X. Han, The Born--Infeld vortices induced from a generalized Higgs mechanism, {\em Proc. Roy. Soc.} A {\bf 475} (2016) 20160012.

\bibitem{FW}
F. Wilczek, Magnetic Flux, Angular Momentum, and Statistics, {\em Phys. Rev. Lett.} {\bf 48} (1982) 1144.

\bibitem{DJFA}
D.P. Arovas, J.R. Schrieffer, F. Wilczek and A. Zee, Statistical Mechanics of Anyons, {\em  Nucl. Phys.}
B {\bf 251} (1985) 117.

\bibitem{DHA}
D.C. Tsui, H.L. Stormer and A.C. Gossard, Two-dimensional magnetotransport in the extreme
quantum limit, {\em Phys. Rev. Lett.} {\bf 48} (1982) 1559-1562.

\bibitem{YFEB}
Y.-H. Chen, F. Wilczek, E. Witten and B.I. Halperin, On Anyon Superconductivity, {\em Int. J. Mod.
Phys.} B {\bf 3} (1989) 1001.

\bibitem{Schroers}
B.J. Schroers, The spectrum of Bogomol'nyi solitons in gauged linear sigma models, {\em Nucl. Phys.} B {\bf475} (1996) 440--468.

\bibitem{HY0}
X. Han and Y. Yang,
Magnetic impurity inspired Abelian Higgs vortices,
{\em J. High Energy Phys.} {\bf 2} (2016) 046.

\bibitem{DK}
D. Tong and K. Wong, Vortices and Impurities, {\em J. High Energy Phys.} 01 (2014) 090.


\bibitem{Y1}
Y. Yang, The relativistic non-Abelian Chern-Simons equations,
{\em Commun. Math. Phys.} {\bf 186} (1997) 199.

\bibitem{XCY}
X. Han, C.-S. Lin and Y. Yang, Resolution of Chern-Simons-Higgs Vortex Equations, {\em Commun. Math. Phys}, {\bf 343} (2016) 701-724. 

\bibitem{JT}
A. Jaffe and C. H. Taubes, {\em Vortices and Monopoles}, Birkh\"{a}user,  Boston, 1980.

\bibitem{WY}
S. Wang and Y. Yang, Abrikosov's vortices in the critical coupling, {\em SIAM J. Math. Anal.} {\bf 23} (1992) 1125-1140.


\bibitem{Hooft}
G. 't Hooft, A property of electric and magnetic flux in non-Abelian gauge theories, Nucl. Phys. B, {\bf 153} (1979) 141--160.

\bibitem{Ab}
A. A. Abrikosov, On the magnetic properties of superconductors of the second group, {\em Sov. Phys. JETP} {\bf5} (1957) 1174--1182.


\bibitem{Aubin}
T. Aubin, {\em Nonlinear Analysis on Manifolds: Monge--Amp\'{e}re Equations}, Springer, Berline and New York, 1982.

\bibitem{HHY}
X. Han, G, Haung, Y. Yang, Coexisting vortices and antivortices generated by dually gauged harmonic maps, {\em J. Math. Phys} {\bf 62}
(2021) 103503.

\bibitem{SSY}
L. Sibner, R. Sibner, and Y. Yang, Abelian gauge theory on Riemann surfaces and new topological
invariants, {\em Proc. Roy. Soc.}  A {\bf456} (2000) 593--613.

\bibitem{XY}
A. Xu and Y. Yang, Bogomol'nyi equations and coexistence of vortices and antivortices in generalized Abelian Higgs theories, {\em Proc. Roy. Soc.} A {\bf481} (2025) 20250424.

\bibitem{NM}
N. Manton, Five vortex equations, {\em J. Phys. A} {\bf 50} (2017) 125403.

\bibitem{SBG}
S. B. Gudnason, Nineteen vortex equations and integrability, {\em J. Phys. A} 55 (2022).
405401

\end{thebibliography}
\end{document}